\documentclass[letterpaper, 10 pt, conference]{amsart}

 \usepackage[foot]{amsaddr}
\usepackage{graphicx,dblfloatfix,subcaption}
\usepackage{wrapfig}
\usepackage[innercaption]{sidecap}
\usepackage{tikz,epstopdf,pgfplots,pdfpages}
\usepackage{framed}  
\usepackage{blkarray}
\usepackage{longtable}
\usepackage{enumerate}%
\usepackage[author={Sofie},draft]{pdfcomment}
\usepackage{amsmath,amssymb,amsfonts}
 	\usepackage{algorithm}
	\usepackage{algorithmic}
\usepackage{mathtools}
\usetikzlibrary{decorations.shapes,shapes.geometric,shadows,arrows,automata,positioning,calendar,mindmap,backgrounds,scopes,chains,er,patterns,pgfplots.groupplots}
  \usetikzlibrary{calc} 

\newcommand{\err}{\varepsilon}
 \newcommand{\mathscr}[1]{\mathcal{#1}}
  \newcommand{\mc}[1]{\mathcal{#1}}
	\newtheorem{theorem}{Theorem}
	\newtheorem{thm}[theorem]{Theorem}
	
	\newtheorem{cor}[theorem]{Corollary}
	\theoremstyle{definition}
	\newtheorem{definition}{Definition}
	\newtheorem{defn}[definition]{Definition}
\theoremstyle{example}
	\newtheorem{example}{Example}
			\newtheorem{condition}{Condition}
			\newtheorem{claim}{Claim}
			
			\theoremstyle{remark}
			\newtheorem{remark}{Remark}

 \newcommand{\newsymb}[4]{
 \newcommand{#2}{#3} 
  }

\newsymb{Hilbert}	{\Hilbert}	{\mathcal{H}_2}
{The Hilbert space} 
\newsymb{Reals}		{\Real}		{\mathbb R}
{The set of reals}        
\newsymb{Naturals}	{\Nat}		{\mathbf{N}}
{The set of natural numbers}             
\newsymb{Complex}	{\Complex}	{\mathbb C}  
{The set of complex numbers}

{} 

\newcommand{\po}{\mathbb{P}}     
\newcommand{\p}[1]{\po\left(#1\right)}     

\newcommand{\norm}[1]{\left\Vert#1\right\Vert}
\newsymb{norm}{\normempty}{\norm{\cdot}}  
{The Euclidean norm $\|x\|=\sqrt{x^Tx}.$}

\newcommand{\eps}{\epsilon}  
 
\newsymb{system}{\s}{\mathbf{S}}{The true `(cyber-)physical' system}
\newsymb{model}{\M}{\mathbf{M}}{A mathematical model}
\newsymb{detmodel}{\Md}{\bar{\mathbf{M}}}{The minimal realisation of $\M$ when the noise input is neglected.}

\newsymb{statespace}{\X}{\mathbb{X}}{The state space, i.e. the set of states}     
	\newsymb{state}{\xv}{\mathbf{x}}{An element of the state space $\X$}   
\newsymb{actionspace}{\A}{\mathbb{U}}{The set of controllable inputs}    
	\newsymb{actions}{\acv}{u}{An element of the set of controllable input $\U$}   
\newsymb{outputspace}{\Y}{\mathbb{Y}}{The set of output values}     
    \newsymb{output}{\yv}{y}{An element of the set of outputs}
\newsymb{performanceoutput}{\Z}{\mathbb{Z}}{The set of performance outputs}     

\newsymb{observer}{\Obs}{\mathbf{O}}{An observer}

\newsymb{TS}{\TS}{\Sigma}{A transition system}

	\newsymb{XTS}{\XTS}{\mathscr{X}}{The set of states of a transition system}
	\newsymb{xTS}{\xTS}{x}{The set of states of a transition system}
	\newsymb{UTS}{\UTS}{\mathscr{A}}{The set of inputs to a transition system} 
	\newsymb{uTS}{\uTS}{u}{The set of inputs to a transition system}
	\newsymb{ZTS}{\ZTS}{\mathscr{Z}}{The set of possible infinite observations}
	 \newsymb{HTS}{\HTS}{\mathscr{H}}{The labelling map}
	
\newsymb{initialstates}{\init}{\mathbb{I}}{The set of initial states}

\newsymb{Markov}{\Markov}{\mathcal{M}}{A Markov process}
\newsymb{Relation}{\rel}{\mathcal{R}}{A relation}
\newsymb{EquivRelation}{\erel}{\mathcal{R}}{An equivalence  relation}

\newsymb{similated}{\simu}{\preceq_{\mathcal{S}}}{Simulated by}
\newsymb{bisimilar}{\bsimu}{\sim_{\mathcal{B}}}{Bisimulation equivalence}
\newsymb{simulationequiv}{\esimu}{\simeq_{\mathcal{S}}}{Simulation equivalence}
\newsymb{altsimilated_app}{\asimu}{\preceq^\eps_{\mathcal{AS}}}{$\eps$-approximate alternating simulation}
\newsymb{similated_app}{\simuapp}{\preceq^\eps_{\mathcal{S}}}{$\eps$-approximate simulated by}

\newsymb{until}{\un}{\mbox{\textbf{U}}}{until operator}
\newcommand{\always}{\Box}
\newcommand{\eventually}{\Diamond}

\newcommand{\InF}{\mathcal{U}_{v}}

\newcommand{\C}{{\mathbf{C}}}

\makeatletter
\newcommand\tabfill[1]{%
\dimen@\linewidth%
\advance\dimen@\@totalleftmargin%
\advance\dimen@-\dimen\@curtab%
\parbox[t]\dimen@{#1\ifhmode\strut\fi}%
}
\newcommand{\Wt}{\mathbb{W}_{\mathbb{T}}}
\newcommand{\pcm}[2]{\po_{\scalebox{0.5}[.5]{$\!\!#1\!\!\times\!\!#2$}}}

\title[approximate similarity relations and policy refinement]{Verification of general Markov decision processes by approximate similarity relations and policy refinement}

\author[S. Haesaert]{S. Haesaert\textsuperscript{1}}
\address{\textsuperscript{1}Department of Electrical Engineering, 
Eindhoven University of Technology}
\author[S. Esmaeil Zadeh Soudjani ]{S. Esmaeil Zadeh Soudjani\textsuperscript{2}}
\author[A. Abate]{A. Abate\textsuperscript{2}}
\address{\textsuperscript{2}Department of Computer Science, 
University of Oxford }
 
\begin{document}
\begin{abstract}
In this work we introduce new approximate similarity relations that are shown to be key for policy (or control) synthesis over general Markov decision processes.  
The models of interest are discrete-time Markov decision processes, 
endowed with uncountably-infinite state spaces and metric output (or observation) spaces. 
The new relations, underpinned by the use of metrics, allow in particular for a useful trade-off between   
deviations over probability distributions on states, 
and distances between model outputs.  
We show that the new probabilistic similarity relations, 
inspired by a notion of simulation developed for finite-state models,   
can be effectively employed  over general Markov decision processes for verification purposes, 
and specifically for control refinement from abstract models.    
\end{abstract}

\maketitle         
\section{Introduction}
  \setcounter{footnote}{1}
The formal verification of computer systems allows for the quantification of their properties and for their correct functioning.  
Whilst verification has classically focused on finite-state models, 
with the ever more ubiquitous embedding of digital components into physical systems  
richer models are needed and correct functioning can only be expressed over the combined behaviour of both the digital computer and the surrounding physical system.  
It is in particular of interest to synthesise the part of the computer software that controls or interacts with he physical system automatically, 
with low likelihood of malfunctioning. Furthermore, 
when computers interact with physical systems such as biological processes, power networks, and smart-grids, 
stochastic models are key. 
Consider, as an example, a power network for which we would like to quantify the likelihood of blackouts and to synthesise strategies to minimise this.

Systems with uncertainty and non-determinism can be naturally modelled as Markov decision processes (MDP).  
In this work, we focus on general Markov decision processes (gMDP) that have uncountable state spaces as well as metric output spaces. 
The characterisation of properties over such processes cannot in general be attained analytically \cite{Abate1}, 
so an alternative is to approximate these models by simpler processes that are prone to be mathematically analysed or algorithmically verified \cite{SAID}, %
such as finite-state MDP \cite{FAUST13}.   
Clearly, it is then key to provide formal guarantees on this approximation step,  such that solutions of the verification or synthesis problem for a property on the simpler process can be extended to the original model. %
Our verification problems include the synthesis of a policy (or a control strategy) that maximises the likelihood of the specification of interest. %

In this work we develop a new notion of approximate similarity relation, 
aimed to attain a computationally efficient controller synthesis over Markov decision processes with metric output spaces. 
We show that it is possible to obtain a control strategy for a gMDP as a refinement of a strategy synthesised for an abstract model, 
at the expense of accuracy defined on a similarity relation between them, 
which quantifies bounded deviations in transition probabilities and output distances. 
In summary, 
we provide results allowing us to quantitatively relate the outcome of verification problems performed over the simpler (abstract) model to the original (concrete) model, 
and further to refine control strategies synthesised over the abstract model to strategies for the original model.

\smallskip 

The use of similarity relations on \emph{finite-state} probabilistic models has been broadly investigated, 
either via exact notions of probabilistic simulation and bisimulation relations  
\cite{larsen1991bisimulation,Segala1995,Segala1995a},   
or (more recently) via approximate notions \cite{Desharnais2008,cDAK12}. 
On the other hand, 
similar notions over \emph{general, uncountable-state spaces} have been only recently studied:  
available relations either hinge on stability requirements on model outputs \cite{Julius2009a,ZMMAL14} (established via martingale theory or contractivity analysis), 
or alternatively enforce structural abstractions of a model \cite{desharnais2004metrics} by exploiting continuity conditions on its probability laws \cite{Abate2011,bcAKNP14}.  

In this work,   
we want to quantify properties with a certified precision \emph{both} in the deviation of the probability laws for finite-time events (as in the classical notion of probabilistic bisimulation) and of the output trajectories (as studied for dynamical models).  
Additionally, we impose no strict requirements on the dynamics of the given gMDP and its abstraction. 
To these ends, 
we first extend the exact probabilistic simulation and bisimulation relations based on lifting for finite-state probabilistic automata and stochastic games \cite{Segala1995,Segala1995a,Zhang2010a} to gMDP (Section \ref{sec:exact}).  
We then generalise these notions to allow for errors on the probability laws \emph{and} deviations over the output space (Section \ref{sec:epsdelta}). 
Two case studies in the area of smart buildings (Section \ref{sec:case}) are used to evaluate these new approximate probabilistic simulation relations. 
Unlike cognate recent work \cite{Abate2011,Julius2009a}, 
we are interested in similarity relations that allow refining over the concrete model a control strategy synthesised on the abstract one. 
We zoom in on relations that, quite like the alternating notions in \cite{Alur1998,Tabuada2009b} for non-probabilistic models and in \cite{Zhang2010a} for stochastic ones, 
quantitatively bound the difference in the controllable behaviour of pairs of models (namely a gMDP and its abstraction).  
In Appendix \ref{sec:lit} we show how over a class 
of Markov processes (without controls), 
this newly developed approximate similarity relation practically generalises notions of probabilistic \mbox{(bi-)simulations} of Labeled Markov processes \cite[based on zigzag-morphisms]{Desharnais2002},\cite[based on equivalence relations]{Desharnais2003}, and  
 their  approximate versions \cite[based on binary relations]{desharnais2004metrics,Desharnais2008,cDAK12}.

\section{Verification of general Markov decision processes: problem setup}  
\subsection{Preliminaries and notations} 
Given two sets $A$ and $B$,  the Cartesian product of $A$ and $B$ is given as $A\times B=\{(a,b): a\in A \textmd{ and } b \in B\}$. 
The disjoint union of $A$ and $B$ is denoted as $A\sqcup B$ and consists of the combination of the members of $A$ and $B$, where the original set membership is the distinguishing characteristic that forces the union to be disjoint,i.e., 
\(A\sqcup B=(A\times \{0\})\bigcup (B\times \{1\}).\) As usual for $C\subset A\sqcup B$ we denote $C\cap A=\{a\in A: (a,0) \in C\}$.
For the sets $A$ and $B$ a relation $\rel\subset A\times B$ is a subset of their Cartesian product that relates elements $x\in A$ with elements $y\in B$, denoted as $x\rel y$.
We use the following notation for the mappings $\rel(\tilde A):=\{y: x\rel y,\  x\in \tilde A\}$ and $\rel^{-1}(\tilde B):=\{x: x\rel y,\  y\in \tilde B\}$  for $\tilde A\subseteq A$ and $\tilde B\subseteq B$.
A relation over a set 
 defines a preorder 
  if it is reflexive, $\forall x\in A: x\rel x$; and transitive, $\forall x,y,z\in A:$ if $x \rel y $ and $y \rel z$ then $x\rel z$. 
A relation $\rel\subseteq A\times A$ is an equivalence relation if it is reflexive, transitive and symmetric, $\forall x,y\in A:$ if $x \rel y $ then $y\rel x$.
 
A measurable space is a pair $(\X,\mathcal{ F})$ with sample space $\X$ and $\sigma$-algebra $\mathcal{F}$ defined over $\X$, 
which is equipped with a topology.    
As a specific instance of $\mathcal F$ consider the Borel measurable space $(\X,\mathcal{B}(\X))$. 
In this work, we restrict our attention to Polish spaces and generally consider the Borel $\sigma$-field \cite{bogachev2007measure}. 
Recall that a Polish space is a separable completely metrisable topological space. 
In other words, 
the space admits a topological isomorphism to a complete metric space which is dense with respect to a countable subset. 
A simple example of such a space is the real line.  

A probability measure $\p{\cdot}$ for $(\X,\mathcal{ F})$ is a non-negative map, 
$\p{\cdot}:\mathcal{ F}\rightarrow [0,1]$ such that $\p{\X}=1$ and such that for all countable collections $\{A_i\}_{i=1}^\infty$ of pairwise disjoint sets in $\mathcal{F}$, 
it holds that  
$\p{\bigcup_i A_i }=\sum _i \p{A_i}$.   
Together with the measurable space, such a probability measure $\po$ defines the probability space, which is denoted as $(\X,\mathcal F,\po)$ and has realisations  $x\sim \po$.   
Let us further denote the set of all probability measures for a given measurable pair $(\X,\mathcal{ F})$ as $\mathcal P (\X,\mathcal{ F})$.  
\setcounter{footnote}{0} 
For a probability space\footnote{The index $\X$ in $\mathcal F_\X$ distinguishes the given $\sigma$-algebra on $\X$ from that on $\Y$, which is denoted as $\mathcal F_\Y$. 
Whenever possible this index will be dropped.}  $(\X,\mathcal F_\X,\po)$ and a measurable space $(\Y,\mathcal F_\Y)$, a $(\Y,\mathcal F_\Y)$- valued \emph{random variable}  is a function $y:\X\rightarrow \Y$ that is $(\mathcal F_\X,\mathcal F_\Y)$-measurable, 
 and which induces the probability measure $y_\ast \po$ in $\mathcal P(\Y,\mathcal F_\Y)$.  
For a given set $\X$ a metric or distance function $\mathbf d_\X$ is a function $\mathbf{d}_\X: \X\times \X\rightarrow \mathbb R_0^+$. 

\subsection{gMDP models - syntax and semantics}
General Markov decision processes are related to control Markov processes \cite{Abate2011} and Markov decision processes \cite{bible,mt1993,hll1996}, 
and formalised as follows.  
 \begin{defn}[Markov decision process (MDP)] \label{def:MDP}
The tuple $\M=(\X,\pi,\mathbb T,\A)$ defines a discrete-time MDP over an uncountable state space $\X$, and is characterised by $\mathbb T$, a conditional stochastic kernel that assigns to each point $x\in \X$ and control $u\in \A$ a probability measure $\mathbb T(\cdot\mid x,u)$ over $(\X,\mathcal B(\X))$. For any set $A\in \mathcal{B}(\X)$, $\po_{x,u}(x(t+1)\in A)=\int_A \mathbb T(dy\mid x(t)=x,u)$, where $\po_{x,u}$ denotes the conditional probability $\po(\cdot\mid x,u)$. The initial probability distribution is $\pi:\mathcal{B}(\X)\rightarrow [0,1]$. 
\end{defn}

At every state the state transition depends non-deterministically on the choice of $u\in \A$.  
When chosen according to 
 a distribution  $\mu_u:\mathcal{B}(\A)\rightarrow [0,1]$, we refer to the stochastic control input as $\mu_u$. Moreover   
the transition kernel is denoted as $\mathbb T(\cdot| x, \mu_u)=\int_\A \mathbb T(\cdot| x, u) \mu_u(du)\in \mathcal P(\X,\mathcal B(\X))$. 
Given a string of inputs 
$u(0), u(1), \ldots, u(N)$,  
over a finite time horizon $\{0,1,\ldots, N\}$, 
and an initial condition  $x_0$ (sampled from distribution $\pi$), 
the state at the $(t+1)$-st time instant, $x(t+1)$,
is obtained as a realisation of the controlled Borel-measurable stochastic kernel $\mathbb{T}\left(\cdot\mid x(t), u(t) \right)$ -- 
these semantics induce paths (or executions) of the MDP.   
 \begin{defn}[General Markov decision process (gMDP)] 
$\M\!=\!(\X,\!\pi,\!\mathbb T,\!\A,\!h,\! \Y)$ is a discrete-time gMDP consisting of an MDP combined with output space $\Y$ and a measurable output mapping $h:\X\rightarrow\Y$.   
A metric $\mathbf d_\Y$ decorates the output space $\Y$.  
\end{defn}
The gMDP semantics are directly inherited from those of the MDP. 
Further, output traces of gMDP are obtained as mappings of MDP paths, namely 
$\{y(t)\}_{0:N}:= y(0), y(1), \ldots, y(N)$, 
where $y(t) = h\big(x(t)\big)$. Denote the 
class of all gMDP with the metric output space $\Y$ as $\mathcal{M}_\Y$. 
Note that gMDP 
can be regarded as a super-class of the known labelled Markov processes (LMP) \cite{desharnais2004metrics} as elucidated in \cite{bcAKNP14}. 
%
\begin{example}\label{ex11}
Consider the stochastic process \begin{align*}\textstyle \M:  
x(t+1)=f(x(t),u(t))+ e(t),\hspace{1.5cm} y(t)=h(x(t))\in\Y,
\end{align*}
with variables $ x(t), u(t), e(t)$, 
taking values in $\mathbb R^n$, 
representing the state, control input\footnote{ In other domains one also refers to the control variables as actions (Machine Learning, Stochastic Games) or as external non-determinism (Computer science).}, and noise
terms respectively. 
The process is initialised as $x(0) \sim \pi$, 
and driven by $e(t)$, 
a white noise sequence with zero-mean normal distributions and covariance matrix $\Sigma_e$. 
This stochastic process, defined as a dynamical model, 
is a gMDP characterised by a tuple $(\mathbb R^n,\pi,\mathbb T,\mathbb R^n, h, \Y)$, 
where the conditional transition kernel is defined as $\mathbb T(\cdot\mid x,u)=\mathcal{N}\left(f(x(t),u(t)),\Sigma_e\right)$, 
a normal probability distribution with mean $f(x(t),u(t))$ and covariance matrix $\Sigma_e$.\qed
\end{example} 

A policy is a selection of control inputs based on the past history of states and controls. 
We allow controls to be selected via universally measurable maps \cite{bible} from the state to the control space, 
so that time-bounded properties such as safety can be maximised \cite{Abate1}. 
When the selected controls are only dependent on the current states, 
and thus conditionally independent of history (or memoryless), 
the policy is referred to as Markov. 
\begin{defn}[Markov policy]\label{def:markovpolicy}
For a gMDP $\M=(\X,\pi,\mathbb T,\A, h, \Y)$, a Markov policy $\mu$ is a sequence $\mu=(\mu_1,\mu_2,\mu_3,\ldots)$ of universally measurable maps $\mu_t=\X\rightarrow \mathcal P(\A,\mathcal B(\A))$ $t=0,1,2,\ldots$, from the state space $\X$ to the set of controls.
\end{defn}
Recall that a function $f:\Z_1\rightarrow \Z_2$ is universally measurable  if the inverse image of every Borel set is measurable with respect to every complete probability measure on $\Z_1$ that measures all Borel subsets of $\Z_1$. 

The execution
$\{x(t), t\in[0,N]\}$ initialised by $x_0\in\X$ and controlled with Markov policy $\mu$ is a stochastic process defined on the canonical sample space $\Omega:= \X^{N+1}$ endowed with its product topology $\mathcal B (\Omega)$. 
This stochastic process has a probability measure $\po$ uniquely defined by the transition kernel $\mathbb T$, policy $\mu$, 
and initial distribution $\pi$ \cite[Prop. 7.45]{bible}.  \\
Of interest are time-dependent properties such as those expressed as specifications in a temporal logic of choice.   
This leads to problems where one maximises the probability that a sequence of labelled sets is reached within a time limit and in the right order. 
One can intuitively realise that in general the optimal policy leading to the maximal probability is not a Markov (memoryless) policy, as introduced in Def. \ref{def:markovpolicy}. 
We introduce the notion of a control strategy, and define it as a broader, memory-dependent version of the Markov policy above. 
This strategy is formulated as a Markov process that takes as an input the state of the to-be-controlled gMDP. 
\begin{defn}[Control strategy]\label{def:CS}
A control strategy $\C=(\X_\C,x_{\C0},\X,\mathbb T^t_\C,h_\C^t)$ for a gMDP $\M$ with state space $\X$ and control space $\A$ 
 over the time horizon $t=0,1,2,\ldots,N$ is an \emph{inhomogenous Markov process} with 
state space $\X_\C$;  
an initial state $x_{\C0}$;  inputs  $x\in\X$; 
time-dependent, universally measurable kernels $\mathbb T^t_{\C}$, $t=0,1,\ldots,N$;  
and with universally measurable output maps $h^t_\C:\X_\C\rightarrow \mathcal P(\A,\mathcal B(\A))$, $t=1,\ldots,N$,
with elements $\mu\in \mathcal P(\A,\mathcal B(\A))$. \qed  
\end{defn}
Unlike a Markov policy, the control strategy is in general dependent on the history, as it has an internal state that can be used to remember relevant past events.  
As elucidated in Algorithm \ref{alg:CM}, 
note that the first control $u(0)$ is selected by drawing $x_\C(1)$ according to $\mathbb T^0_\C(\,\cdot\,{\mid} x_\C(0),x(0))$, 
where $x_\C(0) = x_{\C0},$ and selecting $u(0)$ from measure $\mu^0_{\C}=h_\C^0(x_\C(1))$.\footnote{Note that the stochastic transitions for the control strategy and the gMDP are selected in an alternating fashion. The output map of the strategy is indexed based on the time instant at which the resulting policy will be applied to the gMDP. }
The control strategy applied to $\M$ can be both stochastic (as a realisation of \mbox{$\mathbb T^0_\C(\cdot\,{\mid}\, x_\C(0),x(0))$\,}), 
a function of the initial state $x(0)$, 
and of time.  

The execution $\{(x(t), x_\C(t)), t\in[0,N]\}$ of a gMDP $\M$ controlled with strategy $\C$ 
is defined on the canonical sample space $\Omega:= (\X\times\X_\C)^{N+1}$ endowed with its product topology $\mathcal B (\Omega)$. 
This stochastic process is associated to a unique probability measure $\pcm{\C}{\M}$, 
since the stochastic kernels $\mathbb T^t_\C$ for $t\in[0,N]$ and $\mathbb T$ are Borel measurable and composed via universally measurable policies \cite[Prop. 7.45]{bible}. 
%
\begin{algorithm}[htp]
\caption{Execution of the controlled model $\mathbf C\times \M$}\label{alg:CM}
\begin{algorithmic}
\STATE{set $t:=0$ and $x_\C (0):=x_{\C 0}$ }
\STATE{draw $x(0) \sim \pi$ }\COMMENT{from $\M$}
\WHILE{$t<N$}
\STATE{draw  $x_\C(t+1) \sim \mathbb T^t_\C(\,\cdot\,{\mid} x_\C(t),x(t) )$ }\COMMENT{from $\C$}
\STATE{set $\mu_t:=h_\C^t(x_\C(t+1))$, draw $u(t)$ from $\mu_t$}
\STATE{draw $x(t+1) \sim \mathbb T(\,\cdot\,{\mid}x(t),u(t) ))$ }\COMMENT{from $\M$}
\STATE{set $t:=t+1$}
\ENDWHILE
\end{algorithmic}
\end{algorithm}
 
\subsection{gMDP verification and strategy refinement: problem statement} 
We qualitatively introduce the main problem that we want to solve in this work: 
How can one provide a general framework to synthesise control policies over a formal abstraction $\tilde\M$ of a concrete complex model $\M$,  
with the understanding that $\tilde\M$ is much simpler to be manipulated (analytically or computationally) than $\M$ is? 
We approach this problem by defining a simulation relation under which a policy $\tilde{\mathbf{C}} $ 
for the abstract Markov process $\tilde\M$ implies the existence of a policy  $ {\mathbf{C}} $ for $\M$, 
so that we can quantify differences in the stochastic transition kernels and in the output trajectories for the two controlled models.  
This allows us to derive bounds on the probability of satisfaction of a specification for $\M \times {\mathbf{C}}$ from 
the satisfaction probability of modified specifications for $\tilde\M \times \tilde{\mathbf{C}}$. %
We will show that with this setup we can deal with finite-horizon temporal properties, including safety verification as a relevant instance.

The  results in this paper are to be used in parallel with optimisation, 
 both for selecting the control refinement and for synthesising a policy on the abstract model. 
It has been shown in \cite{bible} that stochastic optimal control  even for a system on a ``basic"  space can lead to measurability issues: 
in order to avoid these issues we follow \cite{bible,Desharnais2008} and the developed theory for Polish spaces and Borel (or universally) measurable notions.  Throughout the paper we will give as clarifying examples Markov processes evolving, as in Example \ref{ex11}, over Euclidean spaces which are a special instances of Polish spaces. 
 This allows us to elucidate the theory. 
 
\section{Exact (bi-)simulation relations based on lifting}\label{sec:exact}
\subsection{Introduction}
In this section we define probabilistic simulation and bisimulation relations that are, respectively, a preorder and an equivalence relation on $\mathcal{M}_\Y$. 
Before introducing these relations, 
we first extend Segala's notion \cite{Segala1995,Segala1995a} of \emph{lifting} to uncountable state spaces, which allows us to equate the transition kernels of two given gMDPs.  
Thereafter, we leverage liftings to define (bi-)simulation relations over $\mathcal{M}_\Y$, which characterise the similarity in the controllable behaviours of the two gMDPs.  
Subsequently we show that these similarity relations also imply controller refinement, 
i.e., within the similarity relation a control strategy for a given gMDP can be refined to a controller for another gMDP. 
In the next section, we show that this exact notion of similarity allows a more general notion of approximate probabilistic simulation. 
The new notions of similarity relations extend the known exact notions in \cite{larsen1991bisimulation}, and the approximate notions of \cite{Desharnais2008,cDAK12}. 
Additionally,  we will show that these results can be naturally extended to allow for both differences in probability and deviations in the outputs of the two gMDPs. 
%

We work with pairs of gMDP put in a relationship, denoting them with numerical indices ($\M_1, \M_2$), 
with the intention to apply the developed notions to an abstraction $\tilde \M$ of a concrete model $\M$, 
respectively.  

\subsection{Lifting for general Markov decision processes}
 
Consider two gMDP $\M_1,\M_2\in \mathcal{M}_\Y$ mapping to a common output space $\Y$ with metric $\mathbf d_\Y$.
For $\M_1=(\X_1,\pi_1,\mathbb T_1,\A_1, h_1, \Y)$ and $\M_2=(\X_2,\pi_2,\mathbb T_2,\A_2, h_2, \Y)$ at given state-action pairs $x_1\in \X_1,u_1\in \A_1$ and $x_2\in \X_2,u_2\in \A_2$, respectively, 
we want to relate the corresponding transition kernels, 
namely the probability measures $\mathbb T_1(\cdot\mid x_1,u_1)\in\mathcal{P}(\X_1,\mathcal{B}(\X_1))$  and $\mathbb T_2(\cdot\mid x_2,u_2)\in\mathcal{P}(\X_2,\mathcal{B}(\X_2))$. 

Similar to the coupling of measures in $\mathcal{P}(\X,\mathcal F)$ \cite{art2014,lindvall2002lectures}, 
consider the \emph{coupling} of two arbitrary probability spaces $(\X_1,\mathcal F_1,\po_1)$ and $(\X_2,\mathcal F_2,\po_2)$ (cf. \cite{skala1993,strassen1965}).  
A probability measure $\po_c$ 
defined on $(\X_1\times \X_2, \mathcal F)$ \emph{couples} the two spaces if the projections $p_1$, $p_2$,  
with $x_1=p_1(x_1,x_2)$ and $x_2=p_2(x_1,x_2)$, 
define respectively an $(\X_1,\mathcal F_1)$- and an $(\X_2,\mathcal F_2)$-valued random variables, 
such that $\po_1=p_{1\ast}\po_c$ and $\po_2=p_{2\ast} \po_c$.   
For \emph{finite- or countable-state} stochastic processes a related concept has been introduced in \cite{Segala1995,Segala1995a}  
and referred to as \emph{lifting}:  
the transition probabilities are coupled using a weight function in a way that respects a given relation over the combined state spaces. %
Rather than using weight functions over a countable or finite domain \cite{Segala1995}, 
we introduce lifting as a coupling of measures over Polish space and their corresponding Borel measurable $\sigma$-fields. 

Since we assume that the state spaces are Polish and have a corresponding Borel $\sigma$-field for the given probability spaces $(\X_1,\mathcal B(\X_1),\mathbb{P}_1)$ and $(\X_2,\mathcal B(\X_2),\mathbb{P}_2)$ with 
  $\mathbb{P}_1:=\mathbb T_1(\cdot\mid x_1,u_1)$  and $\mathbb{P}_2:=\mathbb T_2(\cdot\mid x_2,u_2)$, the natural choice for the $\sigma$-algebra becomes  $\mathcal B(\X_1\times \X_2)=\mathcal{B}(\X_1)\otimes\mathcal{B}(\X_2)$ \footnote{ $\mathcal B(\X_1) \otimes\mathcal{B}( \X_2)$ denotes the product $\sigma$-algebra of $\mathcal B(\X_1)$ and $\mathcal{B}( \X_2)$.}  and the question of finding a coupling can be reduced to finding a probability measure in  $\mathcal P(\X_1\times\X_2,\mathcal B(\X_1\times \X_2))$.  
 
\begin{figure}[htp] 
     \centering
     \includegraphics[width=3in]{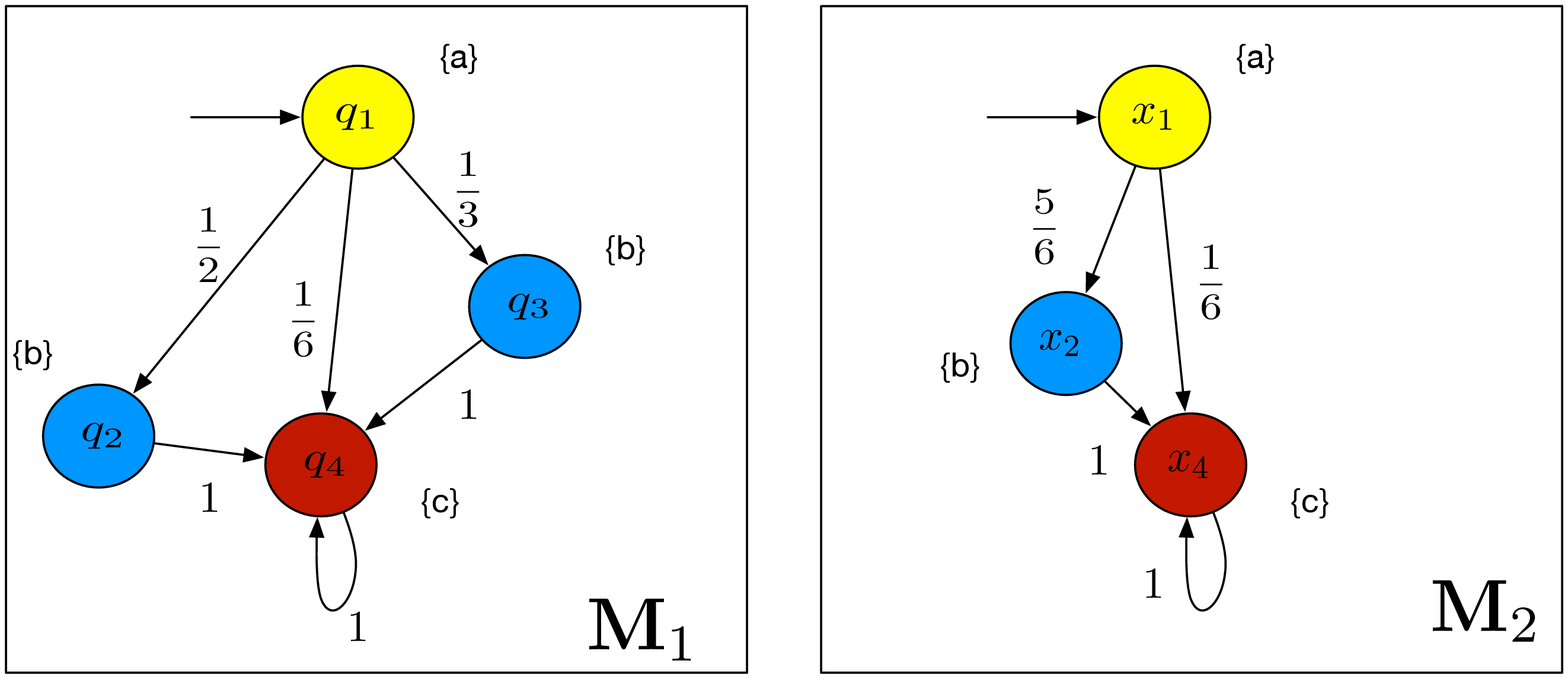} 
     \includegraphics[width=1.75in]{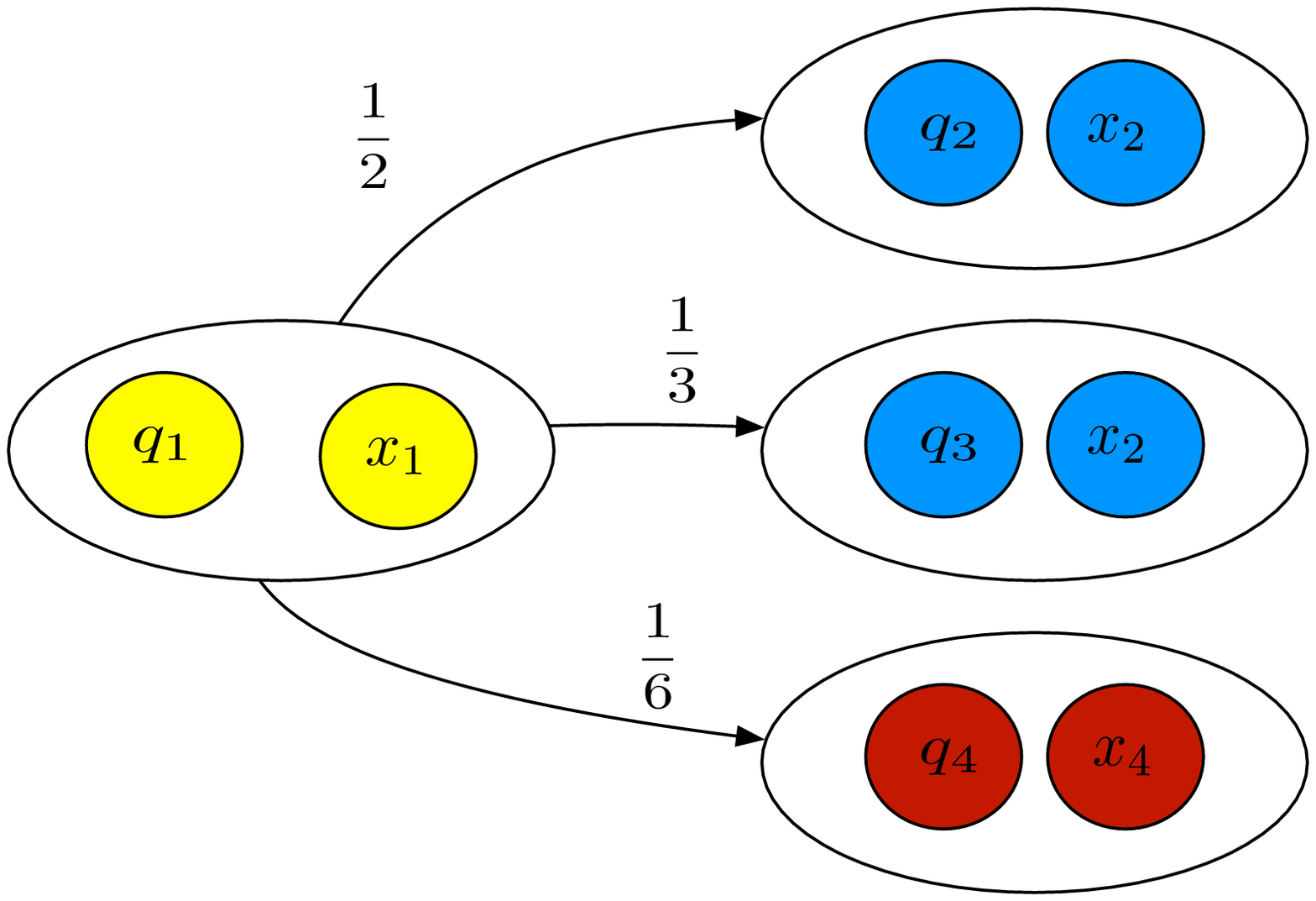} 
     \caption{ Finite-state Markov processes $\M_1$ and $\M_2$ (left $\&$ middle) with  $S=\{q_1,q_2,q_3,q_4\}$ and $T=\{x_1,x_2,x_4\}$ the respective state spaces.  
   The states are labelled with three different colours.
   Lifting probabilities of the transition kernels for $(q_1,x_1)$ are given on the edges of the rightmost figure.}
     \label{fig:lifting1}
  \end{figure}

\begin{defn}[Lifting for general state spaces] \label{def:lifting}
Let $\X_1,\X_2$ be two sets with associated measurable spaces $(\X_1,\mathcal B(\X_1))$ and $(\X_2,\mathcal B(\X_2))$ and let the Borel measureable set $\rel\subseteq \X_1\times \X_2$ be a relation. 
We denote by 
$\bar\rel\subseteq \mathcal{P}(\X_1,\mathcal B(\X_1))\times \mathcal{P}(\X_2,\mathcal B(\X_2))$ the corresponding lifted relation, 
so that $\Delta \bar \rel \Theta$ holds if there exists a probability space $(\X_1\times \X_2,\mathcal B(\X_1\times \X_2)
, \mathbb W)$ 
(equivalently, a lifting $\mathbb W$) satisfying 
{ \setlength{\parskip}{-1pt}\setlength{\parsep}{0pt} \begin{enumerate}
\item for all $X_1\in \mathcal{B}(\X_1)$: $\mathbb W(X_1\times \X_2)=\Delta(X_1)$;
\item  for all $X_2\in \mathcal{B}(\X_2)$:  $\mathbb W(\X_1\times X_2)=\Theta(X_2)$;
\item for the probability space  $(\X_1\times \X_2,\mathcal B(\X_1\times \X_2), \mathbb W)$ it holds that 
$x_1\rel x_2$ with probability $1$, or equivalently that $\mathbb{W}\left(\rel\right)=1$.
\end{enumerate}}\noindent
\end{defn} 
\smallskip 
With reference to the connection with the notion of coupling, 
an equivalent definition of lifting is obtained be replacing $1.$ and $2.$ by the condition that for $(\X_1\times \X_2,\mathcal B(\X_1\times \X_2), \mathbb W)$  the projections $p_1$, $p_2$, with $x_1=p_1(x_1,x_2)$ and $x_2=p_2(x_1,x_2)$, we can define $(\X_1,\mathcal B(\X_1))$ and $(\X_2,\mathcal B(\X_2))$-valued random variables $\Delta=p_{1\ast} \mathbb W$ and $\Theta=p_{2\ast} \mathbb W $. 
An example is portrayed in Fig. \ref{fig:lifting1} containing two models $\M_1, \M_2$ and a relation (denoted by equally labelled/coloured pairs of states), where the transition kernels for a pair of states is lifted with respect to the relation. 

\begin{remark}\label{rem:measdiag}
Notice that the extension of the notion of lifting to general spaces has required the use of \emph{measures}, 
rather than weight functions over a countable or finite domain, as in \cite{Segala1995}. 
We have required that the $\sigma$-algebra $\mathcal B(\X_1\times \X_2)$ contains not only sets of the form $X_1\times \X_2$ and $\X_1\times X_2$, 
but also specifically the sets that characterise the relation $\rel$. 
Since the spaces $\X_1$ and $\X_2$ have been assumed to be Polish, 
it holds that every open (closed) set in $\X_1\times \X_2$ belongs to $\mathcal B(\X_1) \otimes\mathcal{B}( \X_2)= \mathcal B(\X_1\times \X_2)$ \cite[Lemma 6.4.2]{bogachev2007measure}. 
As an instance 
consider the diagonal relation \(\rel_{diag}:=\{(x,x):x\in \X\}\) over $\X\times\X$, 
of importance for examples 
 introduced later. 
This 
is a Borel measurable set 
 \cite[Theorem 6.5.7]{bogachev2007measure}. 
\qed
\end{remark}

\subsection{Exact probabilistic (bi-)simulation relations via lifting} 
  
Similar to the alternating notions for probabilistic game structures in \cite{Zhang2010a}, 
we provide a simulation that relates any input chosen for the first process with one for the second process. 
As such, we allow for more elaborate handling of the inputs than in the probabilistic simulation relations discussed in \cite{Desharnais2008,cDAK12}, 
and further pave the way towards the inclusion of output maps.  
We extend the notions in \cite{Segala1995,Zhang2010a} by allowing for more general Polish spaces. 
Further, we introduce the notion of \emph{interface function} in order 
to connect the controllable behaviour of two gMDP:     
\begin{align*}\InF: \A_1\times \X_1\times\X_2 \rightarrow \mathcal{P}(\A_2,\mathcal B(\A_2)),\end{align*}
where we require 
that $\InF$ is a Borel measurable function. 
This means that $\InF$ induces a Borel measurable stochastic kernel, 
again denoted by $\InF$, 
over $\A_2$ given $(u_1,x_1,x_2)\in\A_1\times \X_1\times\X_2$.
The notion of interface function is known in the context of correct-by-design controller synthesis and of hierarchical controller refinement \cite{Girard2009,Tabuada2009b}.   
For the objective of hierarchical controller refinement, 
an interface function implements (or refines) any control action synthesised over the abstract model to an action for the concrete model. 
In order to establish an exact simulation relation between abstract and concrete models,  
we can attempt to refine the control actions from one model to the other by choosing an interface function that matches their stochastic behaviours.  
On the other hand in the next section, the interface function will be used to establish approximate simulation relations: 
for this goal, the optimal selection of the interface function is the one that optimises the accuracy of the relation.  
This is topic of ongoing research. 

In this work we extend standard interface functions for deterministic systems by allowing randomised actions $\mu_2\in\mathcal P(\A_2, \mathcal B (\A_2))$. 
The lifting of the transition kernels for the chosen interface generates a stochastic kernel $\Wt$ conditional on the values of signals in $\A_1$ and in $\X_1\times\X_2$.  
Let us trivially extend the interface function to \(\InF(\mu_1,x_1,x_2):=\int_{\A_1}\InF(u_1,x_1,x_2)\mu_1(du_1).\) 
\begin{defn}[Probabilistic simulation]\label{def:pbsim}
Consider two gMDP $\M_i, i=1,2$, $\M_i = (\X_i,\pi_i ,\mathbb T_i,\A_i,h_i,\Y)$.    
The gMDP $\M_1$ is stochastically simulated by $\M_2$ if there exists an interface function $\InF$ and
a relation $\rel\subseteq \X_1\times \X_2\in\mathcal B(\X_1\times \X_2)$, for which there exists a Borel measurable stochastic kernel $\Wt(\,\cdot\,{\mid} u_1,x_1,x_2)$ on $\X_1\times\X_2$ given $\A_1\times\X_1\times\X_2$,
such that { \setlength{\parskip}{-1pt}\setlength{\parsep}{0pt}
\begin{enumerate}
\item $\forall (x_1,x_2)\in \rel$, $ h_1(x_1)=h_2(x_2)$;  
\item $\forall (x_1,x_2)\in \rel$, $\forall {u_1}\in \A_1,$  
\(\mathbb T_1(\cdot| x_1,u_1)\ \bar \rel\  \mathbb T_2(\cdot| x_1,\InF(u_1,x_1,x_2)),\) with lifted probability measure $\Wt(\,\cdot\,{\mid} u_1,x_1,x_2)$;  
\item $\pi_1\bar\rel \pi_2$.
\end{enumerate} }%
\noindent The relationship between the two models is denoted as $\M_1\preceq\M_2$.  
\end{defn}  
The Borel measurability for both $\InF$ (see above) and $\Wt$ (as in this definition), 
which is technically needed for the well posedness of the controller refinement, 
can be relaxed to universal measurability, as will be discussed in the Appendix.  
\begin{defn}[Probabilistic bisimulation]\label{def:pbbi}
Under the same conditions as above, 
$\M_1$ is a probabilistic bisimulation of $\M_2$ if there exists a   relation $\rel\subseteq \X_1\times \X_2$ such that $\M_1\preceq \M_2$ w.r.t. $\rel$ and $\M_2\preceq\M_1$ w.r.t. the inverse relation $\rel^{-1}\subseteq \X_2\times \X_1$.
$\M_1$ and $\M_2$ are said to be probabilistically bisimilar, 
which is denoted $\M_1\approx\M_2$.   
\end{defn}
 For every gMDP $\M$: 
\(\M\preceq\M \textmd{ and } \M\approx\M\).
This can be seen by considering the diagonal relation $\rel_{diag}=\{(x_1,x_2)\in\X\times \X\mid x_1=x_2\}$ 
and selecting equal inputs  for the associated interfaces. 
The resulting equal
 transition kernels  $\mathbb{T}(\cdot|x,{u})\bar\rel_{diag}\mathbb{T}(\cdot|x,{u})$
  are lifted by the measure $\Wt(dx_1'\times dx_2'{\mid} u, x_1,x_2)=\delta_{x_1'}(dx_2')\mathbb{T}(dx_1'|x_1,u)$  where $\delta_{x_1'}$ denotes the Dirac distribution located at $x_1'$.  
\begin{example}[%
Lifting for diagonal relations]\label{ex:1} \mbox{ }\\ 
\noindent{$\bf a.$}~Consider the gMDP $(\M_1)$ introduced in Ex. \ref{ex11} and a slight variation of it $(\M_2)$, given as stochastic dynamic processes, 
\begin{align*}\textstyle&\M_1:  
x(t+1)=f(x(t),u(t))+ e(t),&\ y(t)&=h(x(t)),&\\ \textstyle&\M_2:   
x(t+1)=f(x(t),u(t))+ \tilde e(t)+\tilde u(t), &\ 
y(t)&=h(x(t)), &
\end{align*}
with variables $ x(t),x(t+1), u(t), \tilde u(t) , e(t),\tilde e(t)$ taking values in $\mathbb R^n$, 
and with dynamics initialised with the same probability distribution at $t=0$ and driven by white noise sequences $e(t),\tilde e(t)$, 
both with zero mean normal distributions and with variance $\Sigma_e, \Sigma_{\tilde e}$, respectively.  
Notice that if $\Sigma_e-\Sigma_{\tilde e}$ is positive definite then $\M_1\preceq \M_2$.  To see this, 
select the control input pair $(u_2,\tilde u_2) \in \A_2$ as $u_2=u_1$, 
and $\tilde u_2$ according to the zero-mean normal distribution with variance $\Sigma_e-\Sigma_{\tilde e}$, 
then the associated \emph{interface }is $\InF(\,\cdot\,{\mid}u_1,x_1,x_2)=\delta_{u_1}(du_2)\mathcal N(d\tilde u_2{\mid} 0,\Sigma_e-\Sigma_{\tilde e})$.  
For this interface the stochastic dynamics of the two processes are equal, 
and can be lifted with $\rel_{diag}$.\\*
\noindent{$\bf b.$}~
Similar as above, consider 
two gMDP modelled as Gaussian processes 
\begin{align*}&\M_1:  
x(t+1)=(A+BK)x(t)+ Bu(t)+ e(t),&\ y(t)&=h(x(t)),&\\ &\M_2:   
x(t+1)=Ax(t)+B u(t)+ e(t), &\ 
y(t)&=h(x(t)), &
\end{align*} 
with variables $ x(t),x(t+1), e(t)$ taking values in $\mathbb R^n$ and $u(t)\in \mathbb R^m$, matrices $A\in \mathbb R^{n\times n}$, $B\in \mathbb{R}^{n\times m}$, $K\in\mathbb R^{m\times n}$.  Then  $\M_1\preceq\M_2$, since in $\rel_{diag}$
 for every action $u_1$ chosen for $\M_1$, the choice of 
\emph{interface}  $u_2=u_1+Kx_2$ for $\M_2$ results in the same transition kernel for the second model.

\end{example}
\begin{remark} 
Over $\mathcal M_\Y$, 
the class of gMDP with a shared output space, 
the relation $\preceq$ is a preorder, 
since it is reflexive (see Example \ref{ex:1}) and transitive (see later Cor.~\ref{cor:prop}). 
Moreover the relation $\approx$ is an equivalence relation as it is also symmetric (see Cor.~\ref{def:pbbi}). 
\end{remark} 
\subsection{Controller refinement via probabilistic simulation relations}\label{sec:refine_exact}

The ideas underlying the controller refinement are first discussed,  
after which it is shown that the refined controller induces a strategy as per Def. \ref{def:CS}. 
Finally the equivalence of properties defined over the controlled gMDPs is shown. 

Consider two gMDP $\M_i=(\X_i,\pi_i,\mathbb T_i,\A_i,h_i,\Y)$ $i=1,2$ with $\M_1\preceq \M_2$.
Given the entities $\InF$ and $\Wt$ associated to 
$\M_1\preceq \M_2$, 
the distribution of the next state $x'_2$ of $\M_2$ is given as $\mathbb T_2(\cdot\mid  x_2,\InF(u_1,x_1,x_2))$, 
and is equivalently defined via the lifted measure as the marginal of $\Wt(\cdot{\mid} u_1, x_1, x_2)$ on $\X_2$. 
Therefore, the distribution of the combined next state $(x'_1,x'_2)$, 
defined as $\Wt(\,\cdot\,{\mid}u_1,x_1, x_2)$, 
can be expressed as
\begin{align*} 
\Wt(dx_1'\times d x_2'{\mid}  u_1, x_1, x_2)= 
\Wt(d x_1'{\mid} x_2',u_1, x_1, x_2) 
\mathbb T_2(dx'_2{\mid} x_2,\InF( u_1, x_1, x_2)), 
\end{align*}
where $ \Wt(d x'_1{\mid} x'_2,  u_1,  x_1, x_2)$ is referred to as the 
conditional probability given $x'_2$ (c.f. \cite[Corollary 3.1.2]{borkar2012probability}).\footnote{ Beyond Borel measurability, this also holds when the kernels are universally measurable, as corresponding universally measurable regular conditional probability measures are obtained \cite{Edalat1999a}.} 
Similarly, the conditional measure for the initialisation $\mathbb W_\pi$  
 is denoted as 
$\mathbb W_\pi(d x_{1}(0)\times d x_2(0))=\mathbb W_\pi(d  x_1(0){\mid} x_2(0)) \pi_2 (d x_2(0))$. 

Now suppose that we have a control strategy for $ \M_1$, referred to as ${\mathbf{C}}_1$, and 
we want to construct the refined control strategy $\mathbf{C}_2$ for $\M_2$,  
which is such that  
events defined over the output space have equal probability. 
This refinement procedure follows directly from the interface and the conditional probability distributions, 
and is described in Algorithm \ref{alg:refinement}. 
This execution algorithm is separated into the refined control strategy $\C_2$ and its gMDP $\M_2$. 
$\C_2$ is composed of $\C_1$,  the stochastic kernel $\Wt$, and the interface $\InF$, 
and it remembers the previous state of $\M_2$ (cf. line \ref{alg2:remx2} in Algorithm \ref{alg:refinement}). \smallskip

\begin{algorithm}[htp]\caption{Refinement of control strategy ${\mathbf{C}}_1$ as $\mathbf{C}_2$}\label{alg:refinement}\begin{algorithmic}[1]
\STATE{set $t:=0$\\
\STATE draw  $x_2(0)$ from $\pi_2$, \\
\STATE draw $x_1(0)$ from $\mathbb{W}_\pi(\cdot \mid x_2(0))$.  
}
\LOOP
\STATE{given $x_1(t)$, select $u_1(t)$ according $ {\mathbf{C}}_1$, }
\STATE  set $\mu_{2t}:=\InF(u_1(t),x_1(t),x_2(t))$, \\*
\STATE draw $x_2(t+1)$ from $\mathbb{T}_2(\,\cdot\mid x_2(t),\mu_{2t})$,
 \STATE  draw $x_1(t+1)$ from $\Wt(\,\cdot\,|x_2(t+1), u_1(t), x_1(t),x_2(t))$,  \label{alg2:remx2}
\STATE  set $t:=t+1$. 
\ENDLOOP
\end{algorithmic} %
\end{algorithm}%
\begin{thm}[Refined control strategy]\label{thm:Cs1}
Let gMDP $\M_1$ and $\M_2$ be related as $\M_1\preceq\M_2$, 
and consider the control strategy $\C_1=(\X_{\C_1},x_{\C_10},\X_1,\mathbb T_{\C_1}^t,h_{\C_1}^t)$ for $ \M_1$ as given. 
Then there exists at least one refined control strategy $\C_2=(\X_{\C_2},x_{\C_20},\X_2,\mathbb T_{\C_2}^t,h_{\C_2}^t)$, 
as defined in Def. \ref{def:CS}, with 
\begin{itemize}
\item state space $\X_{\C_2}:=\X_{\C_1}\times \X_1\times \X_2$, 
with elements $x_{\C_2}=(x_{\C_1}, x_1,x_2)$; 
\item initial state $x_{\C_20}:=(x_{\C_10},0,0)$;
\item input variable $x_2\in\X_2$, namely the state variable of $\M_2$;  
\item time-dependent stochastic kernels $\mathbb T^t_{\C_2}$, defined 
 as
\begin{align*} 
\mathbb T^0_{\C_2} (dx_{\C_2}{\mid}x_{\C_20},x_2(0) )\ \, &:=\mathbb T^0_{\C_1} (dx_{\C_1}{\mid}x_{\C_10},x_1)\mathbb W_\pi(dx_1 {\mid} x_2 )\delta_{x_2(0)}(dx_2) \mbox{ and }\\
\mathbb T^t_{\C_2} (dx_{\C_2}'{\mid}x_{\C_2}(t),x_2(t) )&:=\mathbb T^t_{\C_1} (dx_{\C_1}'{\mid}x_{\C_1},x_1')\\ &\hspace{-.5cm}
\Wt(d x_1' {\mid} x_2',h_{\C_1}^t(x_{\C_1}),x_2, x_1)\delta_{x_2(t)}(dx'_2) \mbox{ for $t\in[1,N]$};
\end{align*}
\item measurable output maps $h^t_{\C_2}(x_{\C_1},\tilde x_1,x_2):= \InF(h^t_{\C_1}(x_{\C_1}),x_1, x_2)$. 
\qed  
\end{itemize}
\end{thm}
Both the time-dependent stochastic kernels $\mathbb T^t_{\C_2}$ and the output maps $h_{\C_2}^t$, for $t\in[0,N]$, are universally measurable, 
since Borel measurable maps are universally measurable and the latter are closed under composition \cite[Ch.7]{bible}.

Since, by the above construction of $\C_2$, 
the output spaces of the controlled systems $\C_1\times\M_1$ and $\C_2\times \M_2$ have equal distribution, 
it follows that measurable events have equal probability, as stated next and proved in the Appendix. 

\begin{theorem}\label{thm:events} 
If $\M_1\preceq \M_2$, then for all control strategies $\C_1$ there exists a control strategy $\C_2$ such that, 
for all measurable events $A\in \mathcal{B}\left(\Y^{N+1}\right)$,  
\begin{align*}
 \po_{\C_1\times\M_1}\left(\{y_1 (t)\}_{0:N}\in A\right)= \po_{\C_2\times\M_2}\left(\{y_2 (t)\}_{0:N}\in A\right).
\end{align*}
\end{theorem}

\section{{New $\epsilon,\delta$-approximate (bi-)simulation relations via lifting}}\label{sec:epsdelta}  
\subsection{Motivation and $\delta$-lifting}
The requirement on an exact simulation relation between two models is evidently restrictive. 
Consider the following example, where two Markov processes have a bounded output deviation.  

\begin{example}[Models with a shared noise source] \label{Ex:delISS} 
Consider an output space $\Y:=\mathbb R^d$, 
with a metric $\mathbf d_\Y(x,y):= \|x-y\|$ 
(the 
 Euclidean norm), 
and two gMDP expressed as noisy dynamic processes: \\[-1.5em]
\begin{align*}&\M_1:   
x_1(t+1)=f(x_1(t),u_1(t))+ e_1(t)
,&\ y_1(t)&=h(x_1(t)),&\\ &\M_2:   
x_2(t+1)=f(x_2(t),u_2(t))+ e_2(t)
, &\ 
y_2(t)&=h(x_2(t)) , &
\end{align*}\\*[-1.5em]
where $f$ and $h$ are both globally Lipschitz, 
satisfying 
$\|f(x_1,u)-f(x_2,u)\|\leq L \|x_1-x_2\|$ for $0<L<1$,  
and in addition $\|h(x_1)-h(x_2)\|\leq H \|x_1-x_2\|$ for an $0<H$ valid for all $x_1,x_2\in\mathbb R^n$ and for all $u$.  %
Suppose that the probability distributions of the random variable $e_1$ and of $e_2$ depend on a shared noise source $\omega$, with $\omega\in\Omega$ and distribution $\mathbb P_{\omega}$,   
and are such that $e_1(t)=g_1(\omega(t))$ and $e_2(t)=g_2(\omega(t))$.  
Assume now that there exists a value $c\in\mathbb R$, such that $\mathbb P_{\omega}\left[\|g_1(\omega )-g_2(\omega)\|<c\,\right]=1$.  
Then for every pair of states $x_1(t)$ and $x_2(t)$ of $\M_1$ and $\M_2$ respectively, the difference between state transitions is bounded as
\(\textstyle\|x_1(t+1)-x_2(t+1)\|\leq L \|x_1(t)-x_2(t)\|+c\)
with probability $1$. 
By induction it can be shown that if $\|x_1(0)-x_2(0)\|\leq \frac{c}{1-L}$,  
then for all $t\geq 0$, 
$\|x_1(t)-x_2(t)\|\leq \frac{c}{1-L}$, 
and \(\|y_1(t)-y_2(t)\|\leq \frac{cH}{1-L}\). \smallskip\\ 
Even though the difference in the output of the two models is bounded by the quantity $\frac{cH}{1-L}$ with probability $1$,  
it is impossible to provide an approximation error using either the method in \cite{Julius2009a} (hinging on stochastic stability assumptions), nor using (approximate) relations as in \cite{Desharnais2008,cDAK12}:  
with the former approach, for the same input sequence $u(t)$ the output trajectories of $\M_1$ and $\M_2$ have bounded difference, but do not converge to each other;  
with the latter approach, the relation defined via a normed difference cannot satisfy the required notion of transitivity. 
\end{example}
\smallskip
 
As mentioned before and highlighted in the previous Ex. \ref{Ex:delISS}, 
we are interested in introducing a new approximate  
version of the notion of probabilistic simulation relation, 
which allows for both
 $\delta$-differences in the stochastic transition kernels, and $\epsilon$-differences in the output trajectories. %
For the former prerequisite, 
we relax the requirements on the lifting in Def. \ref{def:lifting};  
subsequently, we define the resulting approximate (bi-)simulation relation according to the latter prerequisite on the outputs.   
\begin{defn}[$\delta$-lifting for general state spaces]\label{def:del_lifting}
Let $\X_1,\X_2$ be two sets with associated measurable spaces $(\X_1,\mathcal B(\X_1)), (\X_2,\mathcal B(\X_2))$, 
and let  $\rel\subseteq \X_1\times \X_2$ be a relation for which $\rel\in \mathcal B(\X_1\times \X_2)$. 
We denote by 
$\bar\rel_\delta\subseteq \mathcal{P}(\X_1,\mathcal B(\X_1))\times \mathcal{P}(\X_2,\mathcal B(\X_2))$ the corresponding lifted relation (acting on $\Delta \bar \rel_\delta \Theta$), 
if there exists a probability space $(\X_1\times \X_2,\mathcal B(\X_1\times \X_2), \mathbb W)$  satisfying { \setlength{\parskip}{-1pt}\setlength{\parsep}{0pt}
\begin{enumerate}
\item for all $X_1\in \mathcal{B}(\X_1)$: $\mathbb W(X_1\times \X_2)=\Delta(X_1)$; 
\item  for all $X_2\in \mathcal{B}(\X_2)$:  $\mathbb W(\X_1\times X_2)=\Theta(X_2)$; 
\item for the probability space  $(\X_1\times \X_2,\mathcal B(\X_1\times \X_2), \mathbb W)$ it holds that 
$x_1\rel x_2$ with probability at least $1-\delta$, or equivalently that $\mathbb{W}\left(\rel\right)\geq1-\delta$.  
\end{enumerate}}%
 \end{defn}

We leverage Definition \ref{def:del_lifting} to introduce a new approximate similarity relation that encompasses both approximation requirements, 
obtaining the following $\eps,\delta$-approximate probabilistic simulation. 
\begin{defn}[$\eps,\delta$-approximate probabilistic simulation]\label{def:apbsim}
Consider two gMDP $\M_i=(\X_i,\pi_i ,\mathbb T_i,\A_i,h_i,\Y), i =1,2$,  over a shared {metric} output space  $(\Y,\mathbf{d}_\Y)$.  
$\M_1$ is $\epsilon,\delta$-stochastically simulated by $\M_2$ if there exists an interface function $\InF$ and
a relation $\rel\subseteq \X_1\times \X_2$, for which there exists a Borel measurable stochastic kernel $\Wt(\,\cdot\,{\mid} u_1,x_1,x_2)$ on $\X_1\times\X_2$ given $\A_1\times\X_1\times\X_2$,
such that: 
{ \setlength{\parskip}{-2pt}\setlength{\parsep}{-1pt}
\begin{enumerate}
\item $\forall (x_1,x_2)\in \rel$, $ \mathbf{d}_\Y\left(h_1(x_1),h_2(x_2)\right)\leq \epsilon$;
\item $\forall (x_1,x_2)\in \rel$, $\forall u_1\in\A_1$:  
\(\mathbb T_1(\cdot| x_1, u_1)\ \bar \rel_\delta \  \mathbb T_2(\cdot| x_2, \InF(u_1,x_1,x_2)),\) with lifted probability measure $\Wt(\,\cdot\,{\mid}u_1,x_1,x_2)$; 
\item $\pi_1\bar \rel_\delta \pi_2$.
\end{enumerate} }
\noindent The simulation relation is denoted as $\M_1\preceq^{\delta}_\eps\M_2$.  
\end{defn}
\begin{defn}[$\eps,\delta$-approximate probabilistic bisimulation]\label{def:apbbi}
Under the same conditions as before  $\M_1$ is an $\epsilon,\delta$-probabilistic bisimulation of $\M_2$ if there exists 
a relation $\rel \subseteq \X_1\times \X_2$ such that
$\M_1\preceq_\eps^{\delta}\M_2$ w.r.t. $\rel$ and $\M_2\preceq_\eps^{\delta}\M_1$ w.r.t. $\rel^{-1}\subset \X_2\times \X_1$. \\
\noindent$\M_1$ and $\M_2$ are said to be $\eps,\delta$-probabilistically bisimilar, denoted as $\M_1\approx^\delta_\eps\M_2$.  
\end{defn}
In this section we have provided similarity relations quantifying the difference between two Markov processes. 
The end use of the introduced similarity relations is to quantify the probability of events of a gMDP via its abstraction and to refine controllers: 
this is achieved in the next section.  

\subsection{Controller refinement via approximate 
simulation relations}\label{sec:control}

Consider two gMDP $\M_1$ and $\M_2$, 
for which $\M_1$ is the abstraction of the concrete model $\M_2$.  
The following result is an approximate version of Theorem \ref{thm:events},  
and presents the main result of this paper, 
namely the approximate equivalence of properties defined over the  gMDP $\M_1$ and $\M_2$.  

\begin{theorem}\label{thm:cr} 
If $\M_1\preceq_\eps^\delta \M_2$, 
then for all control strategies $ {\mathbf{C}_1}$ there exists a control strategy $\mathbf C_2$ such that, 
for all measurable events $A\subset \Y^{N+1 }$  
\begin{align*} \pcm{\C_1}{\M_1}\!\left(\{ y_1 (t)\}_{_{0:N}}\!\!\in\! A_{_{\scalebox{0.7}[.7]{\mbox{$-\eps$}}}}\right)-\gamma
 \leq \pcm{\C_2}{\M_2}
 \!\left(\{y_2 (t)\}_{_{0:N}}\!\!\in\! A\right)
\leq\pcm{\C_1}{\M_1}\!\left(\{y_1 (t)\}_{_{0:N}}\!\!\in\! A_{\eps}\right)+\gamma, \end{align*} 
with constant $1-\gamma:=(1-\delta)^{N+1}$,  
and with the $\eps$-expansion of $A$ defined as
 \begin{align*}
 \textstyle&A_\eps:=\!\big\{\{y_\eps(t)\}_{0:N}| \exists \{y(t)\}_{0:N}\in A:  \textstyle\max_{{t\in [0,N]}}  \mathbf d_\Y(y_\eps(t),y(t))\leq \eps\big\}\end{align*}
and  similarly the $\eps$-contraction defined as 
 \(A_{-\eps}:= \{\{y(t)\}_{0:N}|\{\{y(t)\}_{0:N}\}_{\eps}\subset A \}\)
where $\{\{y(t)\}_{0:N}\}_{\eps} $ is the point-wise $\eps$-expansion of $\{y(t)\}_{0:N}$.
\end{theorem}

While the details of the proof can be found in the Appendix, 
its key aspect is the existence of 
a refined control strategy $\mathbf C_2$, 
which we detail next.
Given a control strategy $\mathbf{C}_1$ over the time horizon $t\in \{0,\ldots,N\}$, 
there is a control strategy $\mathbf{C}_2$ that refines $\C_1$ over $\M_2$. The control strategy is conceptually given in Algorithm \ref{alg:ref2}.  
Whilst the state $(x_1,x_2)$ of $\C_2$ is in $\rel$,  
the control refinement from $\C_1$ follows in the same way (cf. Alg.\ref{alg:ref2} line \ref{alg:Rstart}-\ref{alg:Rend})  as for the exact case of Sec. \ref{sec:refine_exact}.
Hence, similar to the control refinement for exact probabilistic simulations, 
the \emph{basic ingredients} of $\C_2$ are the states $x_1$ and $x_2$, 
whose stochastic transition to the pair $(x_1',x_2')$ is governed firstly by a point distribution $\delta_{x_2(t)}(dx_2')$  based on the measured state $x_2(t)$ of $\M_2$; and, subsequently, by the  lifted probability measure
\(
\Wt(dx_1'\mid x_2', u_1,x_2,x_1), 
\) 
\emph{conditioned} on $x_2'$.
On the other hand, 
whenever the state $(x_1,x_2)$ leaves $\rel$ the control chosen by strategy $\C_1$ cannot be refined to $\M_2$: 
instead,  
an  
alternative control strategy $\mathbf C_{rec}$ has to be used to control the residual trajectory of $\M_2$.
The choice is of no importance to the result in Theorem \ref{thm:cr}. 
This stage of the execution (cf. Alg. \ref{alg:ref2}  line \ref{alg:Recstart}-\ref{alg:Recend}) referred to as \emph{recovery} makes the choice of the overall control strategy $\C_2$ non-unique. 
In practice we will only synthesise the control strategy over a finite-time. {\renewcommand{\baselinestretch}{.9}
\begin{algorithm}[htp]
\caption{Refinement of $ {\mathbf{C}}_1$ as $\mathbf{C}_2$}\label{alg:ref2}
\begin{algorithmic}[1]
\STATE{set $t:=0$}\COMMENT{Start}
    \STATE{draw $x_2(0)$ from $\pi_2$}
    \STATE{draw $x_1(0)$ from $\mathbb{W}_\pi(\cdot \mid x_2(0))$}
\WHILE[Refine]{$(x_1(t), x_2(t)) \in \rel$ }\label{alg:Rstart}
\STATE{  given $ x_1(t)$, select $u_1(t)$ from ${\mathbf{C}_1}$,}
\STATE{ set input $\mu_{2t}:=\InF(u_1(t),x_1(t),x_2(t))$, }
\STATE{ draw $x_2(t+1)$ from $\mathbb{T}_2(\,\cdot\mid x_2(t),\mu_{2t})$,}
\STATE{  draw $ x_1(t+1)$ from $\Wt(\,\cdot\,|x_2(t+1),u_1(t),x_1(t),x_2(t))$,  }
\STATE{ set $t:=t+1$}\label{alg:Rend}
\ENDWHILE 
\LOOP[Recover]\label{alg:Recstart}
\STATE{given $ x_2(t)$, select $ \mu_t$ (from $ {\mathbf{C}}_{rec}$),
}
\STATE{ draw $x_2(t+1)$ from $\mathbb{T}_2(\,\cdot\mid x_2(t),\mu_t)$, }
\STATE{ set $t:=t+1$}
\ENDLOOP\label{alg:Recend}
\end{algorithmic} 
\end{algorithm}}

By splitting the execution in Algorithm \ref{alg:ref2} into a control strategy and a gMDP $\M_2$, we can again obtain the refined control strategy. 
\begin{thm}[Refined control strategy]\label{thm:Apprxstrat}
Let gMDP $\M_1$ and $\M_2$, with $\M_1\preceq_\eps^\delta\M_2$, and control strategy $\C_1=(\X_{\C_1},x_{\C_10},\X_1,\mathbb T_{\C_1}^t,h_{\C_1}^t)$ for $ \M_1$ be given. 
Then for any given recovery control strategy $\C_{rec}$, 
a refined control strategy, denoted $\C_2=(\X_{\C_2},x_{\C_20},\X_2,\mathbb T_{\C_2}^t,h_{\C_2}^t)$, can be obtained as an \emph{inhomogenous Markov process} with two discrete modes of operation,  
$\{\operatorname{refinement}\}$ and $\{\operatorname{recovery}\}$, based on Algorithm \ref{alg:ref2}. 
\end{thm}
The details of the tuple $(\X_{\C_2},x_{\C_20},\X_2,\mathbb T_{\C_2}^t,h_{\C_2}^t)$ are given in the Appendix,  
 together with the proof of the theorem. 
They follow from Algorithm \ref{alg:ref2}, in a similar way as Theorem \ref{thm:Cs1} follows from Algorithm \ref{alg:refinement}. \medskip
 
\subsection{Examples and properties}

\begin{example}[Models with a shared noise source -- continued from above]\label{ex:3}\mbox{ }\\*
Based on the relation $\rel:=\{ (x_1,x_2):\|x_1-x_2\|\leq \frac{c}{1-L}\}$ it can be shown that $\M_1\approx_{\eps}^{0}\M_2$ with $\eps=\frac{H c}{1-L}$, 
since, firstly, it holds that $\mathbf{d}_\Y(h(x_1)-h(x_2))\leq \eps$ for all $(x_1,x_2)\in \rel$ with $\mathbf{d}_\Y=\|\cdot\|$. 
Additionally, 
for all $(x_1,x_2)\in \rel$ and for any input $u_1$
the selection $u_2=u_1$ is such that $\mathbb T_1(\cdot|x_1,u_1)\bar\rel_0 \mathbb T_2(\cdot|x_2,u_1)$, note that $\bar\rel_0$ is equal to $\bar\rel$ (the lifted relation from $\rel$).
The lifted stochastic kernel is 
\(\textstyle \Wt (dx_1'\times dx_2'| u_1,x_1,x_2):= \int_{\omega}\delta_{f(x_1,u_1)+g_1(\omega)} (dx_1')  \delta_{f(x_2,u)+g_2(\omega)} (dx_2') \mathbb P_\omega(d\omega),\)
this stochastic kernel is Borel measurable if $f(x_1,u_1)+g_1(\omega)$ and $f(x_2,u)+g_2(\omega)$ are  assumed Borel measurable maps. 
Note that the employed identity interface is also Borel measurable.
\end{example}
\begin{example}[Relationship to model with truncated noise]\label{ex:trunc}
Consider the stochastic dynamical process 
\(\M_1:  
x(t+1)=f(x(t),u(t))+ e(t)\) with output mapping \( y(t)=h(x(t)),\) 
operating over the Euclidean state space $\mathbb R^n$, 
and driven by a white noise sequence $e(t)\in \mathbb R^n$ with distribution $\mathbb P_{e}$. 
The output space $y\in \Y\subseteq\mathbb R^d$ is endowed with the Euclidean norm $\mathbf d_\Y=\|\cdot\|$. 
Select a domain $D\subset \mathbb R^n$ so that, at any given time instant $t$,  
$e(t)\in D$ with probability $1-\delta $. 
Then define a truncated white noise sequence $\tilde e(t)$, with distribution $\mathbb{P}_e\left(\cdot \mid D\right)$. The resulting model $\M_2$ driven by $\tilde e(t)$ is 
\(\M_2:  
x(t+1)=f(x(t),u(t))+ \tilde e(t),\) with the same output mapping 
\(y(t)=h(x(t)).\)
We show that $\M_2$ is a $0,\delta$-approximate probabilistic bisimulation of $\M_1$, i.e. $\M_1\approx_0^\delta\M_2$. 
Select $\mathcal R:=\{(x_1,x_2)\textmd{ for }x_1,x_2\in \mathbb R^n|x_1=x_2\}$, and  choose as interface the identity one, i.e., $\InF(u_1,x_1,x_2)=u_1$. 
A viable lifting measure is 
\begin{align}& 
\textstyle\Wt(dx_1'\times dx_2'|u_1,x_1,x_2):=
 \int_{e\in D} \delta_{x_1'} (dx_2')  \delta_{t_1(e)} (dx_1') \po_e(de)\label{eq:ref truncnoise}\\&\textstyle\hspace{4cm}
+ \int_{e\in\mathbb R^n\setminus D} \delta_{t_1(e)} (dx_1')\po_e(de)
\int_{\tilde e}\delta_{t_2(\tilde e)}  (dx_2') \po_e(d\tilde e|D)  \notag 
 \end{align}
 with $t_1(e)=f(x_1,u_1)+e$ and $t_2(\tilde e)=f(x_2,u_1)+\tilde e$.
 \end{example}
\begin{example}[Relationship between noiseless and truncated-noise models]\label{ex:noiseless}
Consider the model with truncated noise $\M_2$ as defined in Ex.\,\ref{ex:trunc}. 
In what sense is $\M_2$ approximated by its noiseless version $\M_3$, namely
\(\M_3: 
x(t+1)=f(x(t),u(t)),\,
y(t)=h(x(t))\)?
Under requirements on the Lipschitz continuity $\|f(x_1,u)-f(x_2,y)\|\leq L\|x_1-x_2\|$ $0< L<1$,  $\|h(x_1)-h(x_2)\|\leq H\|x_1-x_2\|$,  
and on the boundedness of $D$ and of $c=\max_{d\in D}\|d\|$, 
Ex.\,\ref{Ex:delISS} can be leveraged by concluding that $\M_2\approx_{\eps}^0 \M_3$, 
with $\eps=\frac{Hc}{1-L}$.\footnote{ Alternatively, if $\M_2$ with non-deterministic input $\tilde e\in D$ is an $\eps_a$- alternating bisimulation \cite{Tabuada2009b} of $\M_3$ then $\M_2\approx_{\eps_a}^0\M_3$. 
}\qed 
\end{example}
\smallskip

\noindent In Examples \ref{ex:trunc} and \ref{ex:noiseless} we have that $\M_1$ is approximated by $\M_2$, which is subsequently approximated by $\M_3$. 
The following theorem and corollary attain a quantitative answer on the question whether $\M_1$ is approximated by $\M_3$.
 \begin{theorem}[Transitivity of $\preceq_{\eps}^\delta$]\label{thm:prop}
Consider three gMDP $\M_i$, $i=1,2,3$, defined by tuples $(\X_i,\pi_i ,\mathbb T_i,\A_i,h_i,\Y)$. 
If  \begin{itemize}
\item $\M_1$ is $\epsilon_a,\delta_a$-stochastically simulated by $\M_2$, and
\item $\M_2$ is $\epsilon_b,\delta_b$-stochastically simulated by $\M_3$, 
\end{itemize}%
\noindent then
$\M_1$ is $(\epsilon_a+\epsilon_b),(\delta_a+\delta_b)$-stochastically simulated by $\M_3$. 
Equivalently, if 
\begin{align*}\M_1\preceq^{\delta_a}_{\epsilon_a}\M_2 \textmd{ and }\M_2\preceq^{\delta_b}_{\epsilon_b}\M_3, \textmd{ then }\M_1\preceq^{ \cramped{\delta_a+\delta_b}}_{ \cramped{\epsilon_a+\epsilon_b}}\M_3.\end{align*}
 \end{theorem}
Next, as a corollary of this theorem, 
we derive properties of the notion of approximate bisimulation, 
and discuss the transitivity of the (exact) notions of simulation and of bisimulation relation. 
The latter implies that the simulation relation (cf. Def.\ref{def:pbsim}\,) is a preorder, and that the bisimulation relation (cf. Def.\ref{def:pbbi}\,) is an equivalence relation over the category of gMDP $\mathcal M_\Y$.%
\begin{cor}[Transitivity properties]\label{cor:prop}
Following Theorem \ref{thm:prop}, 
 \begin{itemize} 
 \item 
if $\M_1\approx^{\delta_a}_{\epsilon_a}\M_2$ and $\M_2\approx^{\delta_b}_{\epsilon_b}\M_3$, 
then $\M_1\approx^{ \cramped{\delta_a+\delta_b}}_{ \cramped{\epsilon_a+\epsilon_b}}\M_3$, and 
 \item 
if $\M_1\preceq \M_2$ and $\M_2\preceq \M_3$, then $\M_1\preceq\M_3$, and 
 \item 
if $\M_1\approx \M_2$ and $\M_2\approx \M_3$, then $\M_1\approx \M_3$. 
 \end{itemize}%
\end{cor}
Here notice that for $\rel_{13}:=\{(x_1,x_3)|\exists x_2\in \X_2: (x_1,x_2)\in \rel_{12}, (x_2,x_3)\in \rel_{23}\}$ we show that if $\Delta_1\bar \rel_{12\delta_a}\Delta_2$ and  $\Delta_2\bar \rel_{23\delta_b}\Delta_3$, then $\Delta_1\bar \rel_{13(\delta_a+\delta_b)}\Delta_3$, where the used lifting measure $\mathbb W_{\mathbb T}$ is a function of the respective liftings $\mathbb W_{{\mathbb T}12}$ and $\mathbb W_{{\mathbb T}23}$, i.e. for all $x_1,x_3\in\rel_{13}$ $\exists x_2\in \X_2: (x_1,x_2)\in \rel_{12}, (x_2,x_3)\in \rel_{23}$, $\mathbb W_{\mathbb T}$ is given as  \begin{align*}&\textstyle\Wt(dx_1'\times dx_3'|u_1,x_1,x_2)=\\&\qquad \textstyle\int_{\X_2} \mathbb W_{23}(dx_3'|x_2',{\InF}_{12}(u_1,x_1,x_2),x_2,x_3)\mathbb W_{12}(dx_1'\times d x_2'|u_1,x_1,x_2).\end{align*}
Furthermore,  
the interface ${\InF}_{13}$ is  the composition of ${\InF}_{12}$ and ${\InF}_{23}$. The proof of Theorem \ref{thm:prop} and Corollary \ref{cor:prop} can be found in the Appendix. 

\begin{example}[Combination of Examples \ref{ex:trunc} and \ref{ex:noiseless} via Corollary \ref{cor:prop}]\mbox{ }\\*
For the models in Examples \ref{ex:trunc} and \ref{ex:noiseless} we can conclude that 
 \(\M_1\approx_{\eps}^\delta \M_3. \) 
This means that a stochastic system as in $\M_1$ in Ex.  \ref{ex:trunc}  can be approximated via its deterministic counterpart, 
and that the approximation error can be expressed via the probability (i.e. amount of truncation cf. Ex.  \ref{ex:trunc}) and the output error (i.e. Ex.  \ref{ex:noiseless}). This allows for explicit trading off between output deviation and deviation in probability. 
\end{example} 

\section{Case studies}\label{sec:case}
\subsection{Introduction: energy management in smart buildings}
We are interested in developing advanced solutions for the energy management of smart buildings.  
In this work we first describe a simple example with a three-dimensional model of the thermal dynamics in an office building: 
we consider a simple building that is divided in two connected zones, 
each with a radiator affecting the heat exchange in that zone by controlling the water temperature in a boiler. 
With this case study we aim at elucidating the theory of the previous sections. 
In the third subsection we work with a more realistic model of an office building: 
this 5-dimensional model shows how the given approximate similarity relations can be used for the design of controllers that verifiably satisfy properties expressed as quantitative specifications. 

\subsection{First case study}
A model of the temperature dynamics in an office building with two zones to heat \cite{CDC15a,Holub2013} assumes that the temperature fluctuations in the two zones, 
as well as the ambient temperature dynamics, 
can be modelled 
as a Gaussian process 
\begin{align}%
&\M:
x(t+1)=A x(t)+ B  u(t) + Fe(t),\label{eq:casedyn}\hspace{2cm}\ y(t)=\begin{bsmallmatrix}1&0&0\\0&1&0\end{bsmallmatrix}x(t),
\shortintertext{with stable dynamics characterised by matrices}
A&=\begin{bsmallmatrix}
    0.8725  &  0.0625 &   0.0375\\
    0.0625  &  0.8775 &   0.0250\\
         0  &       0 &   0.9900\end{bsmallmatrix}\notag%
,\quad B =\begin{bsmallmatrix*}[l]
    0.0650& 0\\
         0&0.60\\
         0&0
\end{bsmallmatrix*} ,\quad F=\begin{bsmallmatrix*}[r] 0.05&    -0.02  &       0\\
   -0.02   &  0.05    &      0\\
         0    &     0 &   0.1 \end{bsmallmatrix*}, 
\end{align}%
where $x_{1,2}(t)$ are the temperatures in zone 1 and 2, respectively; 
$x_{3}(t)$ is the deviation of the ambient temperature from its mean; 
and $u(t)\in\mathbb{R}^2$ is the control input.   
The disturbance $e(t)$ is a white noise sequence 
with standard Gaussian distributions, for all $t\in\mathbb R^+$. 
The state variables are initiated as $x(0)=[16\ 14\ \mbox{-}5]^T$.  
This stochastic process can be written as a gMDP, 
as detailed in Example \ref{ex11}. 
As the model abstraction, 
we select the controllable and deterministic dynamics of the mean of the state variables, 
and consequently omit the ambient temperature and the additive noise term: 
 \begin{align}
\tilde\M:\left\{\begin{array}{lll}
 \tilde x(t+1)&= \tilde A\tilde x(t)+ \tilde B  \tilde u(t)\in \mathbb R^2, \mbox{ with }&\tilde A:=\begin{bsmallmatrix}
    0.8725  &  0.0625 \\
    0.0625  &  0.8775 \\
\end{bsmallmatrix},\label{eq:casedyn2}\\
\tilde y(t)&=\begin{bsmallmatrix}1&0\\0&1\end{bsmallmatrix}\tilde x(t),  \quad&\tilde B:=\begin{bsmallmatrix}
    0.0650& 0\\
         0&0.60
\end{bsmallmatrix}.
\end{array}\right.
\end{align}
We then obtain that, as intuitive, $\tilde \M\preceq_{\eps}^{\delta}\M$. 
In order to compute specific values of ${\eps}$ and ${\delta}$, 
we select the relation $\rel:=\{(\tilde x,x)\in \mathbb R^2\times \mathbb R^3\mid  \sqrt{(\tilde x_1- x_1)^2+(\tilde x_2-x_2)^2}\leq \eps \}$ 
and the interface function \(\InF(\tilde u,\tilde x,x)=\tilde u+\tilde B^{-1}(\tilde A \tilde x - \bar Ax )\),  
with $ \bar A = \begin{bsmallmatrix}
    0.8725  &  0.0625 &   0.0375\\
    0.0625  &  0.8775 &   0.0250
    \end{bsmallmatrix}$. 
The structure of the interface is arbitrary: 
in the specific instance the interface is selected to optimally correct the difference in room temperatures at the next time step.  
    
\smallskip    
    
\begin{SCfigure}[1.9][h]
\centering
\caption{Trade-off between the output error $\eps$ and the probability error $\delta$ for the $\delta$,$\eps$-approximate probabilistic simulation $\tilde \M\preceq_{\eps}^\delta \M$. We have selected the pair $(\epsilon, \delta)=(0.16,0.073)$ as an ideal trade-off.\vspace{.3cm}}\label{fig:Casefig}
\resizebox{.4\textwidth}{!}{
 
 \definecolor{mycolor1}{rgb}{0.00000,0.44700,0.74100}%
\begin{tikzpicture}

\begin{axis}[%
width=.4\textwidth,
height=.17\textwidth,
scale only axis,
every outer x axis line/.append style={white!15!black},
every x tick label/.append style={font=\color{white!15!black}},y label style={at={(axis description cs:0.08,.5)}},x label style={at={(axis description cs:(.5,0.08)}},
xmin=0,
xmax=.4,
every outer y axis line/.append style={white!15!black},
every y tick label/.append style={font=\color{white!15!black}},
ymin=0,title={$\tilde \M\preceq_\eps^\delta \M$},   title style={at={(.6,.75)}},
ymax=1,xlabel={
$\eps$},ylabel={
 $\delta$},
axis x line*=bottom,
axis y line*=left
]
\addplot [color=mycolor1,solid,forget plot]  table{Case_fig-1.tsv};
  \draw[ultra thin,dashed] (axis cs:\pgfkeysvalueof{/pgfplots/xmin},0.0733696513683829) -- (axis cs:\pgfkeysvalueof{/pgfplots/xmax},0.0733696513683829);
\draw[ultra thin,dashed] (axis cs:0.16,\pgfkeysvalueof{/pgfplots/ymin}) -- (axis cs:0.16,\pgfkeysvalueof{/pgfplots/ymax});
\node[anchor=west, scale=.7] (n1) at (axis cs:0.16,.2){(0.16,0.073)} ;
\end{axis}
\end{tikzpicture}}\mbox{ } 
\end{SCfigure}
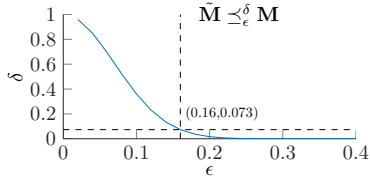

A stochastic kernel $\Wt$ for the lifting is 
\(
\Wt(d\tilde x'\times d x'\mid \tilde u,\tilde x,x ) = \int_{e}
 \delta_{\tilde f }(d\tilde x')\) \(\delta_{f(e)}(d  x')\)\( \mathcal N(de\,{\mid}\,0,I), \)
with $\tilde f =\tilde A\tilde x+\tilde B\tilde u$ and $f(e)=A x+B\InF(\tilde u, \tilde x,x) +F   e$.
The lower bound on $\Wt(\rel\mid \tilde u,\tilde x,x )\leq 1-\delta$ has been computed and traded off against the output deviation, 
as in Fig. \ref{fig:Casefig}. 

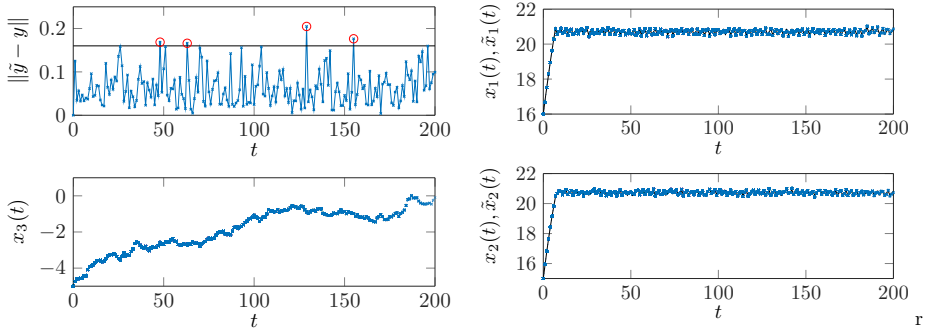
\begin{figure}[htp]
\begin{minipage}{.48\textwidth}\begin{subfigure}{\textwidth}
\resizebox{\textwidth}{!}{\definecolor{mycolor1}{rgb}{0.00000,0.44700,0.74100}%
\begin{tikzpicture}

\begin{axis}[%
width=\textwidth,
height=.3\textwidth,
scale only axis,
separate axis lines,
every outer x axis line/.append style={white!15!black},
every x tick label/.append style={font=\color{white!15!black}},
xmin=0,
xmax=200,
x label style={at={(axis description cs:0.5,0.1)},anchor=north}, 
    y label style={at={(axis description cs:0.1,.6)},anchor=south},
ylabel={$\|\tilde y-y\|$},xlabel={$t$},
every outer y axis line/.append style={white!15!black},
every y tick label/.append style={font=\color{white!15!black}},
ymin=0,
ymax=0.25
]
\addplot [color=mycolor1,solid,mark=x,mark options={solid,scale=.4},forget plot]
  table{normtikz-1.tsv};
\addplot [color=red,only marks,mark=o,mark options={solid},forget plot]
  table{normtikz-2.tsv};
\addplot [color=black,solid,forget plot]
  table{normtikz-3.tsv};
\end{axis}
\end{tikzpicture}}
\end{subfigure}\\[.45em]
\begin{subfigure}{\textwidth}
\resizebox{\textwidth}{!}{\definecolor{mycolor1}{rgb}{0.00000,0.44700,0.74100}%
\begin{tikzpicture}

\begin{axis}[%
width=\textwidth,
height=.3\textwidth,
scale only axis,
x label style={at={(axis description cs:0.5,0.1)},anchor=north},
    y label style={at={(axis description cs:0.1,.6)},anchor=south},
ylabel={$x_3(t)$},
xlabel={$t$},
separate axis lines,
every outer x axis line/.append style={white!15!black},
every x tick label/.append style={font=\color{white!15!black}},
xmin=0,
xmax=200,
every outer y axis line/.append style={white!15!black},
every y tick label/.append style={font=\color{white!15!black}},
ymin=-5,
ymax=1
]
\addplot [color=mycolor1,dotted,mark=x,mark options={solid,scale=.4,opacity=.3},forget plot]
  table{ambienttemp-1.tsv};
\end{axis}
\end{tikzpicture}}
\end{subfigure}\\[-.85em]
\end{minipage}
\begin{subfigure}{.48\textwidth}
\resizebox{\textwidth}{!}{
\definecolor{mycolor2}{rgb}{0.00000,0.44700,0.74100}%
%
\begin{tikzpicture}

\begin{axis}[%
width=\textwidth,
height=.3\textwidth,
x label style={at={(axis description cs:0.5,0.1)},anchor=north},
    y label style={at={(axis description cs:0.1,.6)},anchor=south},
scale only axis, ylabel={$x_1(t),\tilde x_1(t)$},xlabel={$t$},
separate axis lines,
every outer x axis line/.append style={white!15!black},
every x tick label/.append style={font=\color{white!15!black}},
xmin=0,
xmax=200,
every outer y axis line/.append style={white!15!black},
every y tick label/.append style={font=\color{white!15!black}},
ymin=16,
ymax=22,
name=plot1
]
\addplot [color=mycolor2,draw opacity=.75,dotted,mark=x,mark options={solid,scale=.4},forget plot]
  table{tempprof-2.tsv};
\addplot [color=black,draw opacity=1,forget plot]
  table{tempprof-1.tsv};
\end{axis}

\begin{axis}[%
width=\textwidth,
height=.3\textwidth,
x label style={at={(axis description cs:0.5,0.1)},anchor=north},
    y label style={at={(axis description cs:0.1,.6)},anchor=south},
scale only axis,ylabel={$x_2(t),\tilde x_2(t)$},
xlabel={$t$},
separate axis lines,
every outer x axis line/.append style={white!15!black},
every x tick label/.append style={font=\color{white!15!black}},
xmin=0,
xmax=200,
every outer y axis line/.append style={white!15!black},
every y tick label/.append style={font=\color{white!15!black}},
ymin=15,
ymax=22,
at=(plot1.below south west),
anchor=above north west
]

\addplot [color=mycolor2,dotted,mark=x,mark options={solid,scale=.4,opacity=.75},forget plot]
  table{tempprof-4.tsv};
  \addplot[color=black,draw opacity=1,forget plot]
  table{tempprof-3.tsv};
\end{axis}
\end{tikzpicture}%
r}
\end{subfigure}
\caption{Refined  
 control for deterministic model applied to $\M$. The figure (top left) evaluates the accuracy of the approximation, and gives with red circles the instances in which the relation is left. The plot (bottom left) shows the ambient temperature. The plots on the right display the temperature inside the two rooms. The small blue crosses give the actual temperature in the rooms ($x_1,x_2$) whereas the deterministic simulation of ($\tilde x_1,\tilde x_2$) is drawn in black and mostly covered by the crosses. }\label{fig:casecontrol}\vspace{-.6cm}
\end{figure}

We are interested in the goal, expressed for the model $\M$, of increasing the likelihood of trajectories reaching the target set $T = [20.5,\ 21]^2$ and staying there thereafter.
For the abstract model we have developed a strategy, as in\cite{CDC15a},  satisfying by construction the property 
 expressed in LTL-like notation with the formula $\varphi = \eventually \always T$ and shrunken to $\varphi_{-\eps}$ (as per Theorem \ref{thm:cr}).   
This strategy is synthesised as a correct-by-construction controller using \textmd{PESSOA} \cite{Jr2010}, 
where the discrete-time dynamics in \eqref{eq:casedyn2} are further discretised over state and action spaces:  
we have selected 
a state quantisation of $0.05$ over the range $\left[15, 25\right]^2$ for the two state variables, 
and an input quantisation of $0.05$ over the set $\left[10, 30\right]^2$.   
It can be observed that the controller regulates the  abstract model $\tilde \M$ to eventually remain within the target region, as shown in Fig. \ref{fig:casecontrol}. 
We now want to verify that indeed, when refined to the concrete stochastic model, this strategy implies the reaching and staying in the safe set up to some probabilistic error. 
The refined strategy is obtained from this control strategy as discussed in Section \ref{sec:control}, 
and recovers from exits out of the relation $\rel$ by resetting the abstract states in the relation. 

In a simulation study reported in Fig. \ref{fig:casecontrol}, we have executed the refined control strategy over a time horizon of 200 steps. 
Observe that for the execution displayed in the top/left plot  
the behaviour of the controlled concrete model $\M$ remains close to that of $\tilde\M$.  
Only at 4 incidences (circled) does the output error exceed the level $\epsilon = 0.16$. 
This reflects our expectations, 
since at any point in time the probability that the output error exceeds the level $\epsilon = 0.16$ over the following $X$ time steps is provably less than $1-(1-\delta)^X \approx X \delta=0.073X$, as per Theorem \ref{thm:cr}, 
which leads to an upper bound of 15 occurrences. 
Within this case study, whenever the state of the abstract and concrete model leave the relation $\rel$, 
then the recovery strategy consists of resetting the state of the abstract model and continuing with the refined control strategy.  
Thanks to the use of the $\eps$-contraction $\varphi_{-\eps}$ of the concrete specification $\varphi$, 
model $\M$ will still abide by $\varphi$ with a high confidence. 

\subsection{Second case study} 
We consider a realistic model for an office building, with the dynamics obtained from \cite{Bacher2011}.
With a time sampling of $5$ minutes, 
the following model describes stochastic temperature fluctuations around a known mean value: 
\begin{align*}&\mathbf M_{\mbox{office}}:\quad\left\{\begin{array}{ll}
x_b(t+1)&=\Xi x_b(t)+ \Gamma q(t)+B_{p} w_p(t)+ B_s\Phi_s(t)+ B_a T_a(t)\\
y(t)&=\begin{bsmallmatrix}0&1&0&0\end{bsmallmatrix}x_b(t),\end{array}\right.\\
& \begin{bsmallmatrix*}\Xi\mid \Gamma\mid B_p  \mid    B_s\mid B_{a} \end{bsmallmatrix*}= \left[\begin{smallmatrix}  
    0.4487 &    0.216   & 0.2164&    0.1186\\
     0.216 &   0.1778  &  0.3719 &   0.2334\\
   0.09639 &   0.1657 &   0.6569  & 0.08082\\
 0.005234   & 0.0103 & 0.008007    &0.9708\end{smallmatrix}\right|\left.  
 \begin{smallmatrix}    2.65\mbox{e-}5\\
    7.45\mbox{e-}5\\
   2.06\mbox{e-}4\\
    0.07\mbox{e-}5\\
 \end{smallmatrix}\right|\left. 
 \begin{smallmatrix}     1.0939\mbox{e-}4\\
   2.16\mbox{e-}4\\
   7.45\mbox{e-}5\\
   3.92\mbox{e-}6   \end{smallmatrix}\right| \left.\begin{smallmatrix}    6.60\mbox{e-}4\\
   1.31\mbox{e-}3\\
   4.49\mbox{e-}4\\
   2.36\mbox{e-}5
   \end{smallmatrix}\right|\left.
\begin{smallmatrix}
   2.96\mbox{e-}4\\
   8.79\mbox{e-}4\\
   1.93\mbox{e-}4\\
   5.67\mbox{e-}3
 \end{smallmatrix}\right].
%
\end{align*}
The output $y(t)$ models the temperature deviation of the internal air.
The 4-dimensional state of the model, obtained from a frequency-based identification procedure, 
represents the fluctuation of internal temperatures in the building, 
including the building envelope and the interior
\cite[TiTeThTs model]{Bacher2011}, 
where the influence of mean values dynamics have been eliminated from the model. 
The objective of this model is to capture the influence of stochastic effects acting upon the system and control them via the heater with input $q(t)$.
The model represents the stochastic disturbances on the building temperature.  
We foresee three major sources of stochastic disturbance to the system, as explained next. 

The first, $w_p(t)$ is the randomness of the heat generated by people in the building. 
An average person generates $100$ Watt [W] under normal circumstances. 
We presume that the occupancy of the office adds a random element to this average number, 
which we capture as an independently and identically distributed random signal with Gaussian distribution and a standard deviation equal to 20 $\%$ per person:  
when there are $n_p:=10$ people in the office this standard deviation becomes $\sqrt{n_p}\times 20$ [W].
 
The second source of stochastic disturbance is the ambient temperature, 
for which we model the stochastic deviation $T_a(t)$ from accurate weather forecasts. 
As this deviation is correlated over time, this is modelled as a first-order coloured noise,  
with a time constant of 20 minutes. 
The choice of the time constant gives a measure of correlation in time \cite{therrien1992discrete}, 
so we use it to choose the time over which there is a significant correlation between successive values of $T_a(t)$. 
Additionally, we choose it such that the stationary variance is equal to 1, i.e., $\mathbf E \left[T_a(t)^2  \right]=1 $. 
The resulting weather model is a first-order (1-dimensional) model \(T_a(t+1)=0.7788 T_a+0.6273 w_{w}(t)\), 
which is driven by a white noise source with standard Gaussian distribution, namely $w_w(t)\sim\mathcal N(0,I)$.  

The third and final source of disturbance $\Phi_s(t)$ is the energy flow from solar radiation. 
Though measurable, this disturbance cannot exactly be predicted and has a high impact on the temperature inside the office. 
The impact depends on the effective window area of the building, which has been estimated as 6.03 [m$^2$] in \cite{Bacher2011}. 
Based on the measured solar radiation in \cite{Bacher2011}, we model this disturbance as a white noise source with standard deviation of $0.1$ [kW/m$^2$]. 

Including the weather model for $T_a$, which requires encompassing the noise signal $w_{w}(t)$, 
leads to the following 5-dimensional model for the temperature fluctuations in the office building:  
\begin{align*}
&\mathbf M=(A,B,B_w,C):\quad\left\{\begin{array}{ll}
x(t+1)&=Ax(t)+B_w w(t)+ Bu(t)\\
y(t)&=\begin{bsmallmatrix}0&1&0&0&0\end{bsmallmatrix}x(t)\end{array}\right.\\
&\begin{bmatrix}A\mid\! B\mid \!B_w\end{bmatrix}\!=\! \left[\begin{smallmatrix*}[l]  
    0.4487 &   0.216 &  0.2164 &  0.1186 & 2.96\mbox{e-}4 \\ 
   0.216 &  0.1778 &  0.3719 &  0.2334 & 8.789\mbox{e-}4 \\ 
 0.09639 &  0.1657 &  0.6569 & 0.08082 & 1.928\mbox{e-}4 \\ 
0.005234 &  0.0103 &8.007\mbox{e-}3 &  0.9708 &0.005667 \\ 
       0 &       0 &       0 &       0 &  0.7788\end{smallmatrix*}\right|\left.\begin{smallmatrix*}[l]  0.1326 \\ 
  0.3725 \\ 
   1.029 \\ 
4.309\mbox{e-}3 \\ 
       0  \end{smallmatrix*}\right|\left.
 \begin{smallmatrix*}[l]0.006918 & 0.06596 &       0 \\ 
 0.01372 &  0.1308 &       0 \\ 
0.004712 & 0.04492 &       0 \\ 
2.485\mbox{e-}4 &0.002369 &       0 \\ 
       0 &       0 &  0.6273
   \end{smallmatrix*}\right]\!.\end{align*}
In order to avoid numerical ill-conditioning issues, 
both the heat input $q(t)$ (expressed in kW) and the corresponding matrix $\Gamma$ have been replaced by scaled versions, 
namely the input signal $u(t)$ and the input matrix $B$. 
At full throttle the heating input $q(t)=5$[kW] corresponds to the scaled input $u(t)=1$. 
Similarly the three noise sources discussed above have been normalised together with the respective system matrices, 
so that $w(t)$ is the new driving noise, as a white-noise sequence with a standard Gaussian distribution,  
encompassing the unpredicted heat caused by people, solar radiation, and weather fluctuations. 
We are interested in controlling the obtained stochastic system $\mathbf M$ to verify a quantitative property over its output signal, 
which is the inner air temperature. 
More precisely, we want to maximise the probability that the deviation of the inner air temperature stays within a $0.5$ degrees difference from the nominal temperature, 
over an horizon of 30 minutes. 
This property can be encoded as a  PCTL  specification for the discrete time model as follows: 
\(
\mathbb{P}_{\geq p}\left(  \always^6[ |y|<0.5]\right), 
\) where $p$ is a parameter to be optimised over.  

In order to solve this type of probabilistic safety problems we would normally employ formal abstractions, as implemented in the software tool FAUST$^2$ \cite{FAUST13}. 
However, a straightforward use of the tool on the non-autonomous 5-dimensional model does not yield tight guarantees. 
Hence, we first obtain several reduced-order models;  
then, over the input range of interest, we quantify the corresponding $\epsilon,\delta$-approximate probabilistic bisimulation relations;  
finally, we design a controller over the obtained formal abstractions with FAUST$^2$, and refine it to the original 5-dimensional model of the office building.  
In the refinement step we tune the trade-off between the conservativeness with respect to heating inputs and the accuracy of the approximation. 

\subsubsection*{Model abstraction}
We use model order reduction via balanced truncations, as implemented in \texttt{Matlab}, 
to obtain lower-order approximations preserving the dynamics of interest. 
We seek to obtain either first- or second-order models, 
from two types of concrete dynamics: 
firstly, the native dynamics of model $\mathbf M=(A,B,B_w,C)$, 
and secondly the dynamics of model $\mathbf M'=(A+BF,B,B_w,C)$. 
In the latter case, the state-feedback gain $F$ is chosen\footnote{The gain term is obtained with the $\texttt{dare}(A,B,C^TC,0.02)$ command in \texttt{Matlab}. } so that it reduces the importance of the controllable modes of the system: 
\(\
F=\begin{bmatrix}
  0.48456 &     0.39865  &    0.85352   &   0.56387  &  0.0024252
\end{bmatrix}.\)

As a result, we obtain four reduced-order models $\mathbf M_{i}=(A_{i},B_{i},B_{wi},C_{i}) (i=1,2,3,4)$ of $\M$  via balanced truncation\footnote{This results from the application of the 
$\texttt{balred}$ function in \texttt{Matlab}.} : 
 \begin{align}\mathbf M_{i} :\quad\left\{\begin{array}{ll}
x_{s}(t+1)&=A_{i} x_{s}(t)+ B_{wi}w(t)+B_{i} u_{s}(t)\\
y_{s}(t)&= C_{i} x_{s}(t),\end{array}\right.
\end{align}
where the resulting matrices are given in the appendix. 

Models $\mathbf M_{1}$ and $\mathbf M_{3}$ are obtained based on $\mathbf M=(A,B,B_w,C)$, 
whereas  $\mathbf M_{2}$ and $\mathbf M_{4}$ are based on the dynamics of $\mathbf M'=(A+BF,B,B_w,C)$. 
As expected the quality of the reduced models depends on the choice of $\M'$ or $\M$: 
in the former case, the part of the dynamics that we cannot compensate with a control is approximated best,
whereas for $\mathbf M$ the most prominent dynamics are approximated best,  
notwithstanding how well they can be controlled.

\subsubsection*{Approximate probabilistic simulation relations}
The reduced models $\M_1,$ $\M_2,$ $\M_3,\M_4$ are approximations of $\M$ and it is expected that, even when using an interface function, the error between these reduced models and $\M$ will increase with the input $u_s$. 
Therefore we quantify the performance of $\M_i$ for $i=1,2,3,4$ only over a bounded input set  $\mathbb U_s:= \{u_s\in \mathbb R\mid  u_s^2\leq c_1 \}$.
To choose a relevant $c_1$ suppose we would take constant  $c_1$ of $0.25=0.5^2$, then this would be equal to an allowed deviation of 50 percent of the maximal input for the nominal heat input, which is $5$[kW] for the original system. As we only want to correct the heating with respect to stochastic fluctuations we take the more realistic value for $c_1$ of $0.2^2=0.04$. 

Let us now compute the parameters pair $(\epsilon,\delta)$ establishing the relationship $\mathbf M_i\preceq_\epsilon^\delta \mathbf M$ between reduced-order and concrete models.  
Similarly to the work \cite{Girard2009} on hierarchical control based on model reduction we consider a putative relation between the two state spaces as 
 \begin{align*}
 \rel :=\left\{ (x,x_s)\mid 
 (x-Px_s)^T M (x-Px_s)\leq \epsilon^2
 \right\}, 
 \end{align*}
with properly-sized matrices $M$ and $P$, 
satisfying the Sylvester equation $P A_i = AP+ BQ$, for a choice of $Q$, 
and $C_i=CP$, 
and so that $M-C^TC$ is positive semi-definite, namely $M-C^TC\succeq0$. 
Introduce the interface $\InF:\mathbb U_s\times \X_s\times \X\rightarrow \mathbb U$ as 
\[u=R u_s+Qx_s+K(x-Px_s),\] 
and notice that $\InF$ is a function of both $P$ and $Q$ above, 
alongside the additional design variables $R$ and $K$ (to be further discussed shortly).   
The interface function is chosen to reduce the differences in the observed stochastic behaviours of the two systems.  
It refines any choice of $u_s$ to a control input $u$, as such it implements any control strategy for $\M_i$ to the original model $\M$. 
In this case study we have considered a concrete model that is controllable, linear, time-invariant, and driven by an additive stochastic noise.  
The chosen interface $\InF$, with design variables $Q$, $K$, and $R$, 
fully parameterises the set of possible interfaces that refine controls synthesised over a reduced model that is deterministic, linear, and time-invariant, as suggested in \cite{Girard2009}. 

Let us next focus on the characterisation of the relation $\mathbf M_i\preceq_\epsilon^\delta \mathbf M$. 
\noindent Condition 1 in Definition \ref{def:apbsim}, namely 
$\forall (x,x_s)\in \rel: d_\Y(y(t),y_s(t))\leq \epsilon$, 
holds since $\|y-y_s\|^2=\|Cx-C Px_s\|^2$ and $(x-Px_s)^TC^TC(x-Px_s)\leq  (x-Px_s)^TM(x-Px_s)$, 
and the latter is bounded by $\epsilon^2$ for $(x,x_s)\in \rel$.  

\noindent For condition 2, i.e., 
$\forall (x,x_s)$ and $\forall u_s \in \mathbb U_s$:  
\(\mathbb T_s\left(\cdot\mid x_s,u_s \right)\bar \rel_\delta \mathbb T\left(\cdot\mid x,\InF(u_s,x_s,x) \right),\)
we construct a 
lifted probability measure $\mathbb W_{\mathbb T}(\cdot\mid u_s,x_s,x)$ based on the shared input noise $w(t)$. 
From this lifting measure, 
the original transition kernels can easily be recovered by marginalising over $\X_s$ and over $\X$, respectively, as 
 \(\mathbb T\left(\cdot\mid x,u\right) = \mathcal{N}(\cdot| Ax+B\InF(u_s,x_s,x),B_wB_w^T),\) and \( 
 \mathbb T_s\left(\cdot\mid x_s,u_s\right) = \mathcal{N}(\cdot| A_sx_s+B_su_s,B_{wi}B_{wi}^T).\)
 The last condition requires that, 
 with probability at least $1-\delta$,  
the pair $(x', x_s') \in \rel$ is distributed as $(x', x_s')
 \sim \mathbb W_{\mathbb T}\left(\cdot\mid  u_s,x_s,x\right) $.  This condition can be encoded as:
$\forall w^T w\leq c_w$, $\forall (x,x_s)\in\rel$, $\forall u_s\in \mathbb U_s$ it holds that:  $(x'-x_s')\in \rel.$
Note that the latter can be written as $ (x'- P x_s')^TM (x'- P x_s')\leq \eps^2$, where   
\begin{align} \label{eq:diffRelation}
x'- P x_s'&=(A +BK)(x- P x_s) + (B_w -PB_{wi}) w  +( BR   -PB_s) u_s.
  \end{align}
 The conditions above 
 can be expressed as a single matrix inequality  via the $S$-procedure  \cite{Boyd2004}.
We know that $w\sim \mathcal{N}(0,I)$, $w^Tw$ has a Chi-square distribution with 2 degrees of freedom. 
Thus for a required level of $1-\delta$, we select $c_w$  as  $c_w=\chi_2^{-1}(1-\delta)$ and solve the resulting constraints with respect to $\epsilon$ for given values of $K,P,Q$ and $R$, 
for each of the reduced models $\M_i$ using CVX \cite{cvx}.  Note that $\chi_2^{-1}$ is  the chi-square inverse cumulative distribution function with 2 degrees of freedom.
The gains $K$ and $R$ are selected together with $M$  by alternately optimising their choice. 
The chosen $P$ and $Q$ follow from the Sylvester equation,  
for which additional freedom is used to minimise the influence of $w$ and $u_s$ in \eqref{eq:diffRelation}. 
 
Table \ref{apprxsimulation} provides a number of $\epsilon,\delta$ values, 
derived from the approximate probabilistic simulation relation, for each of the models $\M_i$. 
Notice that for increasing values of $\delta$, $\epsilon$ decreases to a positive lower bound: 
this lower bound is a function of the size of the set $\mathbb U_s$. 
Based on these outcomes, we have decided to proceed with $\M_2$. 

\begin{table}[htp]
\caption{$\eps,\delta$-simulation relation trade-off for the reduced-order models. The table gives for each model and $\delta$ the computed $\eps$. }
\label{apprxsimulation}
\begin{tabular}{c|llllllllll}
 $\delta$& \,\,1 &\,$10^{-\frac{1}{3}}$&\,$10^{-\frac{2}{3}} $& \,  $ 10^{-1}$&\,$10^{-\frac{4}{3}}\!$ &\,$10^{-\frac{5}{3}}$ &\,$ 10^{-2}$&\,$10^{-\frac{7}{3}}$&\,$10^{-\frac{8}{3}} $&\,$10^{- {3}}$\\ \hline
\!\!$\M_1$\!\!& \small 0.1233 \!\!\!\!& \small 0.4803 \!\!\!\!& \small 0.6247 \!\!\!\!& \small 0.7347 \!\!\!\!& \small  0.827 \!\!\!\!& \small 0.9082 \!\!\!\!& \small 0.9816 \!\!\!\!& \small  1.049 \!\!\!\!& \small  1.112 \!\!\!\!& \small  1.171 \\ 
\!\!$\M_2$\!\!& \small 0.01445 \!\!\!\!& \small 0.1037 \!\!\!\!& \small  0.132 \!\!\!\!& \small 0.1534 \!\!\!\!& \small 0.1714 \!\!\!\!& \small 0.1871 \!\!\!\!& {\small{\underline{0.2014}}}\!\!\!\!& \small 0.2145 \!\!\!\!& \small 0.2267 \!\!\!\!& \small 0.2381 \\ 
\!\!$\M_3$\!\!& \small0.05206 \!\!\!\!& \small 0.7612 \!\!\!\!& \small  0.997 \!\!\!\!& \small  1.175 \!\!\!\!& \small  1.325 \!\!\!\!& \small  1.456 \!\!\!\!& \small  1.575 \!\!\!\!& \small  1.684 \!\!\!\!& \small  1.785 \!\!\!\!& \small  1.881 \\ 
\!\!$\M_4$\!\!& \small  0.1839 \!\!\!\!& \small 0.3029 \!\!\!\!& \small 0.3358 \!\!\!\!& \small 0.3604 \!\!\!\!& \small 0.3809 \!\!\!\!& \small 0.3988 \!\!\!\!& \small  0.415 \!\!\!\!& \small 0.4298 \!\!\!\!& \small 0.4435 \!\!\!\!& \small 0.4564  \end{tabular}\end{table}

\subsubsection*{Control synthesis over abstract model $\mathbf M_2$: use of FAUST$^2$}  
For a given choice of $\eps,\delta$ we follow Theorem \ref{thm:cr} and modify the given PCTL property 
\(
\psi:=\mathbb{P}_{\geq p}\left( \always^6[ |y|<0.5]\right)
\)
to obtain   
\(
\psi_{\epsilon,\delta}:=\mathbb{P}_{\geq p+\gamma}\left( \always^6[ |y|<0.5-\eps]\right). 
\)
Here $\gamma$ gives the accumulation of the error in the probability over the time horizon of interest: 
for this case we have $1-\gamma:=(1-\delta)^{6}$, which is $\gamma\approx6\delta$.  
We then apply FAUST$^2$ to obtain a grid-based approximation of the safety probability over the six time steps of the formula (which adds up to 30 minutes in the model), 
with an accuracy of $0.1$.
More precisely, 
we first quantise the input space (this on its own generates an exact simulation), 
then we apply FAUST$^2$ \cite{FAUST13} over the obtained continuous space, finite action model.  
For this work we have optimised the algorithms in FAUST$^2$ to use less memory for models with Gaussian noise:  
by first decoupling the noise by means of a simple state transform, the storage of the discretised probability transitions can be done in a structured and more efficient manner. 
This leads to perform the computations with $2.6\times 10^7$ grid points to attain the desired accuracy of 0.1 (more precisely $0.0983$) with a 2,6 GHz Intel Core i5  with 16 GB memory within less then 20 minutes. 
We finally obtain that the modified safety property is satisfied with probability of at least $0.8412-0.0983 = 0.7429$ for the reduced order model $\M_2$ initialised at zero.

\subsubsection*{Control refinement: simulation results} 
We refine the policy obtained from FAUST$^2$  for the reduced-order model $\M_2$ to the original model $\M$. 
Recall that we expect this refined policy to have a quantifiable safety, expressed via the property $\psi$, 
which is a requirement that the inner air temperature remains within the bound $y_{s} \in [-0.5,0.5]$ of the nominal temperature during the next 30 minutes. 
The safety probability for the concrete model $\M$ initialised at the origin is lower bounded by the computed probability $p= (0.7429-\gamma)=( 0.7429-0.0585) = 0.6844$ 
(this is according to Theorem \ref{thm:cr}). 
 
We empirically validate this result as follows. 
We first initialise the system and the state of the reduced-order model (in the controller)  at the origin. 
Then we perform $10^5$ Monte-Carlo simulations and observe that executions of the reduced-order model remain in the modified safe set 85.81 percent of the time, 
whereas they exit it $14.19$ percent of the time.   
For the same noise sequences,   
the controlled $5$-dimensional model, 
where the control is refined based on the interface introduced before, 
stays in the \emph{original} safe set 99.9 percent of the time, and exits it in $0.10$ percent of the times. 
The concrete model is further seen to stay within the \emph{modified} safe set $86.05$ percent of the times, 
which is much closer to the computed probability for the reduced-order model. 
Notice that these empirical outcomes are expected to be higher than indicated in the error bounds, 
as these bounds are conservative especially when considering states starting in the middle of the relation. 
Similarly, starting at the edge of the modified safe set $y_{s}\in[ 0.2986, -0.2986]$ of the reduced-order model, 
we have considered the initialisation as follows $x_s(0)=\cramped{[ -0.4229\  {-0.2987}]^T}$ and $x(0)=Px_s(0)$, 
where $P$ has been discussed above. 
For this initial state $0.7289$ is the lower bound on the safety probability for the reduced-order model, and $p= 0.6704$ for the full-order model.  
With $10^5$ empirical Monte-Carlo runs, 
we obtain that the reduced-order model stays in the modified safe set $84.30$ percent of the time, 
whereas the concrete model with the refined control policy stays in the safe set in $99.87$ percent of the runs. 
Similar results were obtained upon initialising at other points on the edges of the (modified) safe set, 
or on the edge of the relation.

\section{Conclusions}

In this work we have discussed new and general approximate similarity relations for general control Markov processes, 
and shown that they can be effectively employed for abstraction-based verification goals as well as for controller synthesis and refinement over quantitative specifications.    
The new relations in particular allow for a useful trade-off between the deviations in probability distribution on states and the deviations between model outputs. 
We have extended results on control refinement for deterministic LTI systems to construct interface functions effectively. 
For this and other model classes within the set of gMDPs the algorithmic construction of appropriate interface functions together with the optimal quantification of the $\eps,\delta$-approximate similarity relation is topic of further research. %
Alongside practical applications of the developed notions, 
current research efforts focus on further generalisation of Theorem \ref{thm:cr} to specific quantitative properties expressed via temporal logics. 
We are moreover interested in further expanding our understanding of the properties of similarity relations.

\section*{Acknowledgments}

The authors would like to thanks P.M.J. Van den Hof for technical feedback on this work. 

\bibliographystyle{siamplain}
\bibliography{library}%

\newpage

\appendix

\section{Nomenclature}
\begin{longtable}{p{1cm}p{.8\textwidth}}
$\rel$&   Relation over $\rel \subseteq \X_1\times \X_2$.\\
$\bar \rel$&   Relation over $\bar\rel\subseteq\mathcal P (\X_1,\mathcal B(\X_1))\times \mathcal P (\X_2,\mathcal B (\X_2)$ obtained via lifting from $\rel$, as per Def. \ref{def:lifting}.\\
$\bar \rel_\delta$&   Relation over $\bar\rel\subseteq\mathcal P (\X_1,\mathcal B(\X_1))\times \mathcal P (\X_2,\mathcal B (\X_2)$ obtained via the approximate  lifting with a deviation in probability bounded with $\delta$ obtained from $\rel$, as per Def. \ref{def:del_lifting}.\\
$\equiv_{\rel_{eq}}$& Relation between  two probability spaces $(\X_1,\mathcal B(\X_1))$ and $(\X_2,\mathcal B (\X_2))$  based on the equivalence relation $\rel_{eq}\subseteq(\X_1\sqcup\X_2)\times (\X_1\sqcup\X_2)$, \`a la \cite{Desharnais2003}, as reviewed in Section \ref{sec:lit}.\\
$\equiv_{\rel_{eq}}^\delta$& Approximate relation between  two probability spaces $(\X_1,\mathcal B(\X_1))$ and $(\X_2,\mathcal B (\X_2))$  based on the equivalence relation $\rel_{eq}\subseteq(\X_1\sqcup\X_2)\times (\X_1\sqcup\X_2)$,  \`a la \cite{Abate2011}, as  reviewed in Section \ref{sec:lit}.\\
$\preceq$&	Probabilistic simulation relation, see  Def. \ref{def:pbsim} .\\
$\approx$&	Probabilistic bisimulation relation, see Def. \ref{def:pbbi}.			\\
$\preceq_\eps^\delta$& $\eps,\delta$-approximate probabilistic simulation relation, see Def. \ref{def:apbsim}.\\	
\end{longtable}

\section{Details on Case study and use of FAUST$^2$}The model reduction procedure via balanced truncation\footnote{This is obtained from the application of the $\texttt{balred}$ function in \texttt{Matlab}.} 
yields four reduced-order models $\mathbf M_{i}=(A_{i},B_{i},B_{wi},C_{i})$ $i=1,2,3,4$: 
 \begin{align*}
 \mathbf M_{i} :\quad\left\{\begin{array}{ll}
x_{s}(t+1)&=A_{i} x_{s}(t)+ B_{wi}w(t)+B_{i} u_{s}(t)\\
y_{s}(t)&= C_{i} x_{s}(t),\end{array}\right.
\end{align*}
which are characterised by the following constant matrices 
 \begin{align*}
&\mathbf M_{1}: A_{1}=\begin{bsmallmatrix*}[r]            
      0 & -0.8572 \\ 
       1 &   1.857
   \end{bsmallmatrix*}, 
B_{1}=\begin{bsmallmatrix*}[r]     -0.5343 \\ 
  0.5523 \end{bsmallmatrix*}, 
B_{w1}=\begin{bsmallmatrix*}[r]-5.916\mbox{e-}3 & -0.0564 &  8.62\mbox{e-}3 \\ 
 6.138\mbox{e-}3 & 0.05852 &-6.739\mbox{e-}3 
\end{bsmallmatrix*},
C_{1}= \begin{bsmallmatrix*}[l]      0 &       1 \end{bsmallmatrix*}; \!\!\!\!
\\
&\mathbf M_{2}:
A_{2} =\begin{bsmallmatrix*}[l]             
      0 &-0.05267 \\ 
   0.125 & -0.1081 
   \end{bsmallmatrix*},\,
   B_{2}=\begin{bsmallmatrix*}[l]
    0.8917 \\ 
  0.3725\end{bsmallmatrix*},\,
  B_{w2}=\begin{bsmallmatrix*}[l] 0.01925 &  0.1835 &0.002356 \\ 
 0.01372 &  0.1308 &3.229\mbox{e-}5 
\end{bsmallmatrix*},\,
C_{2}= \begin{bsmallmatrix*}[l]    0 &       1 \end{bsmallmatrix*}; 
\\
&\mathbf M_{3}:\,
A_{3} =\begin{bsmallmatrix}              
  0.9951  \end{bsmallmatrix},\,
  B_{3}=\begin{bsmallmatrix}        0.1194 \end{bsmallmatrix},\,
  B_{w3}=\begin{bsmallmatrix*}[l]
0.001497 & 0.01427 & 0.01467\end{bsmallmatrix*},
C_{3}= \begin{bsmallmatrix}       1\end{bsmallmatrix}; 
\\
&\mathbf M_{4}:\,
A_{4}
=\begin{bsmallmatrix}              
0.1203 
   \end{bsmallmatrix},\,
   B_{4}=\begin{bsmallmatrix}
       0.3829 \end{bsmallmatrix},\,
       B_{w4}=\begin{bsmallmatrix*}[l]
     0.01257 &  0.1198 &0.0002907
\end{bsmallmatrix*},\,
C_{4}= \begin{bsmallmatrix}            1\end{bsmallmatrix}. 
\end{align*}

Models $\mathbf M_{1}$ and $\mathbf M_{3}$ are obtained from $\mathbf M=(A,B,B_w,C)$, 
whereas  $\mathbf M_{2}$ and $\mathbf M_{4}$ are based on the dynamics of $\mathbf M'=(A+BF,B,B_w,C)$. 
We have synthesised $F$ to be $\begin{bsmallmatrix} 0.4846 &  0.3986 &  0.8535 &  0.5639 &0.002425 \end{bsmallmatrix}$. 
As expected the reduced models depend on the choice of $\M'$ or $\M$: 
in the former case, the part of the dynamics that we cannot compensate with a control is approximated best,
whereas for $\mathbf M$ the most prominent dynamics are approximated best.

\subsubsection*{Approximate probabilistic simulation relation}
We quantify the performance of $\M_i$ for $i=1,2,3,4$ only over a bounded input set  $\mathbb U_s:= \{u_s\in \mathbb R\mid  u_s^2\leq c_1 \}$.

Subsequently solving the Sylvester equations for $Q,P$ and $R$,   
tuning a stabilising  interface gain $K$, 
and then using the $S$-procedure as described in \cite{Boyd2004} to compute $\eps, \delta$ and $M$, 
we finally obtain the following matrices for the reduced-order models.
For $\M_1$ we take $R:= 1.403$, and we obtain 
\begin{align*}
Q& :=\begin{bsmallmatrix}-0.08954 &-0.07712\end{bsmallmatrix} ,& \quad K:=\begin{bsmallmatrix}  -0.5717 & -0.4705 & -0.9859 & -0.6213 &-0.002364 \end{bsmallmatrix}, \\
P&:= \begin{bsmallmatrix}   -1.061 & 0.09045 \\ 
       0 &       1 \\ 
  -2.295 & -0.9696 \\ 
   9.064 &   8.775 \\ 
       0 &       0  \end{bsmallmatrix}, &\quad
M:=\begin{bsmallmatrix}   0.4797 &  0.1476 &  0.3298 &  0.1397 &-0.001306 \\ 
  0.1476 &   1.104 &  0.1592 & 0.06704 &-0.00359 \\ 
  0.3298 &  0.1592 &  0.2862 &  0.1207 &-0.001327 \\ 
  0.1397 & 0.06704 &  0.1207 &  0.1744 &0.003174 \\ 
-0.001306 &-0.00359 &-0.001327 &0.003174 &0.003676\end{bsmallmatrix}.
\end{align*} 
Note that the latter is optimised for $\delta=10^{-2}$.\\
\noindent For $\M_2$ we take $R:=   1.004$, and obtain  \begin{align*}
Q& :=\begin{bsmallmatrix}  -1.857 &   1.406 \end{bsmallmatrix},
&\quad  K:=\begin{bsmallmatrix} -0.3553 & -0.2931 &   -0.65 & -0.4739 &-0.002547 \end{bsmallmatrix}, \\
P&:= \begin{bsmallmatrix}  -0.6186 &  0.2348 \\ 
       0 &       1 \\ 
   2.562 &  -2.314 \\ 
-0.009378 &0.001329 \\ 
       0 &       0 \end{bsmallmatrix},& \
M:=\begin{bsmallmatrix}       0.2416 & 0.06342 &  0.3159 &  0.1299 & 0.00106 \\ 
 0.06342 &   1.772 & 0.07267 & 0.02663 &0.0007664 \\ 
  0.3159 & 0.07267 &  0.4191 &  0.1728 &0.001395 \\ 
  0.1299 & 0.02663 &  0.1728 & 0.08168 &0.000351 \\ 
 0.00106 &0.0007664 &0.001395 &0.000351 &0.0001456 \end{bsmallmatrix}.
\end{align*}
Again $M$ is chosen based on the $S$ procedure to optimise $\eps$ for $\delta=10^{-2}$.
For $\M_3$, take $R:= 0.3074$ and obtain 
\begin{align*}
Q& :=-0.0008755,&\quad 
K:=\begin{bsmallmatrix} -0.5796 &  -0.477 & -0.9978 & -0.6265 &-0.00236\end{bsmallmatrix}, \\
P&:= \begin{bsmallmatrix}     1.004 \\ 
       1 \\ 
   1.006 \\ 
  0.9713 \\ 
       0\end{bsmallmatrix}, &\quad 
M:=\begin{bsmallmatrix}    8.584 &  -4.974 &   4.929 &   2.078 &  0.1158 \\ 
  -4.974 &   3.944 &  -3.106 &   -1.31 &-0.05919 \\ 
   4.929 &  -3.106 &   3.917 &   1.653 & 0.06135 \\ 
   2.078 &   -1.31 &   1.653 &  0.7024 & 0.02595 \\ 
  0.1158 &-0.05919 & 0.06135 & 0.02595 & 0.01179  \end{bsmallmatrix}.
\end{align*}
Note that $M$ is chosen based on the $S$-procedure to optimise $\eps$ for $\delta=10^{-2}$.

\noindent For $\M_4$, we take  $R := 0.8996$ and 
\begin{align*}
Q& := -0.6961,&\quad K:=\begin{bsmallmatrix}   -0.5307 & -0.4366 & -0.9241 & -0.5946 &-0.002391\end{bsmallmatrix},\\
P&:= \begin{bsmallmatrix}    -1.191 \\ 
       1 \\ 
   1.242 \\ 
-0.01296 \\ 
       0 \end{bsmallmatrix},&\quad 
M:=\begin{bsmallmatrix}  0.03949 &-0.01465 & 0.06076 & 0.02542 &1.999e-05 \\ 
-0.01465 &   1.788 &  0.1162 & 0.05143 &-0.0005164 \\ 
 0.06076 &  0.1162 &   0.128 & 0.05469 &-2.765e-05 \\ 
 0.02542 & 0.05143 & 0.05469 & 0.04108 &-0.0004062 \\ 
1.999e-05 &-0.0005164 &-2.765e-05 &-0.0004062 &0.0003725 \end{bsmallmatrix}.
\end{align*}

\begin{table}[htp]
\begin{tabular}{c|cccccccccc}
 $\delta$& 1 &$10^{-\frac{1}{3}}$&$10^{-\frac{2}{3}} $&   $ 10^{-1}$& $10^{-\frac{4}{3}}$ & $10^{-\frac{5}{3}}$ & $ 10^{-2}$&$10^{-\frac{7}{3}}$&$10^{-\frac{8}{3}} $&  $10^{- {3}}$\\ \hline
$\M_1$  \!\!& \small   0.1233  \!\!\!\!& \small   0.4803  \!\!\!\!& \small   0.6247  \!\!\!\!& \small   0.7347  \!\!\!\!& \small    0.827  \!\!\!\!& \small   0.9082  \!\!\!\!& \small   0.9816  \!\!\!\!& \small    1.049  \!\!\!\!& \small    1.112  \!\!\!\!& \small    1.171 \\ 
$\M_2$  \!\!& \small   0.01445  \!\!\!\!& \small   0.1037  \!\!\!\!& \small    0.132  \!\!\!\!& \small   0.1534  \!\!\!\!& \small   0.1714  \!\!\!\!& \small   0.1871  \!\!\!\!& \small   0.2014  \!\!\!\!& \small   0.2145  \!\!\!\!& \small   0.2267  \!\!\!\!& \small   0.2381 \\ 
$\M_3$\!\!& \small  0.05206  \!\!\!\!& \small   0.7612  \!\!\!\!& \small    0.997  \!\!\!\!& \small    1.175  \!\!\!\!& \small    1.325  \!\!\!\!& \small    1.456  \!\!\!\!& \small    1.575  \!\!\!\!& \small    1.684  \!\!\!\!& \small    1.785  \!\!\!\!& \small    1.881 \\ 
$\M_4$ \!\!& \small    0.1839  \!\!\!\!& \small   0.3029  \!\!\!\!& \small   0.3358  \!\!\!\!& \small   0.3604  \!\!\!\!& \small   0.3809  \!\!\!\!& \small   0.3988  \!\!\!\!& \small    0.415  \!\!\!\!& \small   0.4298  \!\!\!\!& \small   0.4435  \!\!\!\!& \small   0.4564  \end{tabular}
\caption{Trade-off for parameters $\eps,\delta$ in the simulation relation.}
\end{table}

\subsection{FAUST$^2$ computations on a 2-dimensional model}
For a given $x,u$ pair the probability distribution of the next state is distributed with the following stochastic density kernel
\(t_x(\bar x\mid x,u)\sim \mc N(\cdot;A_ix+B_i u,\Sigma )\), 
where $\Sigma:=B_{w_2}B_{w_2}^T$.

We resort to the algorithms implemented in \cite{FAUST13} to maximise the probability of a stochastic event. 
We set up a stochastic dynamic programming scheme, leading to a final value function providing the probability of the property as 
    \[V_0(x)=\mathbb P\left[ \always^6(|y(t)|\leq 0.5-\err)\right].\]
Define the safe set
\(\mc A:=\mathbb R \times [-0.5+\err,0.5-\err]\subset\X=\mathbb R^2,\)
then the property to be maximised can be written as
    \(V_0(x)=\mathbb P\left[ \always^6 \mc A\right].\)

\subsubsection{The error computation}
Assume there are constants $H_1,H_2$, 
such that 
\begin{align}\label{FAUSTeq1}\int_{\mathbb R^2}|t_x(\bar x\mid x,u)-t_x(\bar x\mid x',u)|d\bar x\leq H_1|x_1'-x_1|+H_2|x_2'-x_2|.\end{align}
This gives a linearly increasing error \(N(H_1\Delta_1+H_2\Delta_2)\), 
where $\Delta_i$ is the grid size in the $i$-th coordinate direction of the state space. 
Let us compute the two constants next. Starting from 
\[t_x(\bar x\mid x,u)=\frac{1}{\sqrt{(2\pi)^2\det (\sigma)}}\exp\left[-\frac{1}{2}\left(\bar x-A_ix-B_iu\right)^T\Sigma^{-1}\left(\bar x-A_ix-B_iu\right)\right],\]
define 
\(m=\begin{bmatrix}m_1\\m_2\end{bmatrix}=A_ix+B_i u\) and $\Sigma^{-1}=\begin{bmatrix}d_{11}&d_{12}\\d_{21}&d_{22}\end{bmatrix}=L^TL$. 
Then
\[t_x(\bar x\mid x,u)=\frac{1}{\sqrt{(2\pi)^2\det (\sigma)}}\exp\left[-\|L\bar x-Lm\|^2\right].\]
Define a change of variables with $v=L\bar x $ $\rightarrow$ $dv=|\det(L)|d\bar x$.
Then the error computation follows from the maximal difference between the probability density distributions \cite{FAUST13} as given in \eqref{FAUSTeq1} and can be rewritten as follows: 
\begin{align*}
&\int_{\mathbb R^2}\left|
\frac{1}{\sqrt{(2\pi)^2\det (\Sigma)}}
\left(
\exp\left[-\frac{1}{2}\|v-Lm\|^2
\right]-\exp\left[-\frac{1}{2}\|v-Lm'\|^2
\right]
\right)
\right|\frac{dv}{\det(L)}. 
\shortintertext{Note that $ \Sigma^{-1}=L^TL$, hence $|\det(L)|=\frac{1}{\sqrt{\det(\Sigma)}}$ and consequently }
&=\int_{\mathbb R^2}\frac{1}{ 2\pi}\left|
\left(
\exp\left[-\frac{1}{2}\|v-Lm\|^2
\right]-\exp\left[-\frac{1}{2}\|v-Lm'\|^2
\right]
\right)
\right| dv.  
\shortintertext{Now we can transform a two-dimensional integral into two one-dimensional integrals:}
&\leq \int_{\mathbb R} \frac{1}{\sqrt{2\pi}}
\left|
\left(
\exp\left[-\frac{1}{2}\|v_1-L_1m_1\|^2
\right]-\exp\left[-\frac{1}{2}\|v_1-L_1m'_1\|^2
\right]
\right)
\right|dv_1\\&+
 \int_{\mathbb R} \frac{1}{\sqrt{2\pi}}
\left|
\left(
\exp\left[-\frac{1}{2}\|v_2-L_2m_2\|^2
\right]-\exp\left[-\frac{1}{2}\|v_2-L_2m'_2\|^2
\right]
\right)
\right|dv_2\\
&\leq \frac{2|L_1 m-L_1m'|}{\sqrt{2\pi}}+\frac{2|L_2 m-L_2m'|}{\sqrt{2\pi}}
\leq \frac{2 }{\sqrt{2\pi}} \left(|L_1 A_i(x-x')|+|L_2 A_i(x-x')|
\right).
\end{align*}
Define 
\(\begin{bmatrix}\bar{a}_{11}&\bar{a}_{12}\\
\bar{a}_{21}&\bar{a}_{22}\end{bmatrix}=LA_i.
\)
Then for \eqref{FAUSTeq1} we have $H_1= \frac{2 }{\sqrt{2\pi}} (|\bar{a}_{11}|+|\bar{a}_{21}|)$, $H_2= \frac{2 }{\sqrt{2\pi}} (|\bar{a}_{12}|+|\bar{a}_{22}|)$


\section{Connections to literature and measurability issues}
\label{sec:lit}

In this section we establish quantitative connections between the notion of approximate similarity that we have introduced for gMDPs and known and established concepts that have been discussed in the literature for processes that are special cases of gMDPs.   

As measurability issues are key in this discussion we would like to first point out that the results in this paper can be extended to analytical spaces with universally measurable kernels. 
When we allow the gMDPs to have universally measurable kernels, we need to show  the existence of a conditional probability measure $\Wt(d x'_1|x'_2,u_1,x_1,x_2)$: 
for this we refer to \cite{Edalat1999a} which discusses the existence of universally measurable regular conditional probabilities.   

\subsection{Early results for Markov chains with finite state spaces}
From the perspective of testing, 
the concept of probabilistic bisimulation has been first introduced in \cite{larsen1991bisimulation}, based on a relational notion, 
and later used to define equivalence between Labelled Markov processes (LMPs) \cite{Desharnais2002}. 
LMPs are different from gMDPs in that transition are not governed by actions but by observable labels, 
and the acceptance of a label (and the consequent transition) defines the behaviour of such a process. 
LMPs are defined over a finite state space $\mathbb S$, 
a set of labels $L$, and stochastic transition kernels $\mathbb T_l:\mathbb S\times\mathbb S\rightarrow [0,1]$ that are finitely indexed by $l\in L$. 
There is a strong relationship between LMPs and standard MDPs with labels \cite{bcAKNP14}, despite their different semantics. 
 \begin{defn}[Probabilistic bisimulation (relational notion)]\mbox{ }\\
Let $T=(\mathbb S,\mathbb P_{l\in L},L)$ be a labelled Markov chain, with $L$ the finite set of labels. 
Then a probabilistic bisimulation $\equiv_p$ is an equivalence on $\mathbb S$ such that, whenever $s\equiv_pt$, the following holds:
\begin{align*}\textstyle\forall l\in L:\forall A\in \mathbb S/\equiv_p, \sum_{s'\in A}\mathbb T_l(s|s')=\sum_{s'\in A}\mathbb T_l(t|s').\end{align*}
Two states $s$ and $t$ are said to be  probabilistically bisimilar ($s\sim_{SL}t$) if the pair $(s,t)$ is contained in a probabilistic bisimulation relation.
\end{defn}
An extension of this definition is used to compare two separate processes by combining their state spaces (as a disjoint union) and defining the probabilistic bisimulation on the obtained extended state space \cite{Desharnais2002}. (More details on this operation is given in the following subsection for continuous state-space models.) 
 
For countable-state probabilistic processes combining probability and non-determinism, 
\cite{Segala1995,Segala1995a} has discussed probabilistic simulations based on a lifting notion -- this has inspired the extension (over more general models) that is elaborated in this work. 
Over finite- or countable-state sets, 
\cite[Lemma 8.2.2]{Segala1995} has shown that lifting coincides with $\rel_{eq}$-equivalence of the corresponding probability distributions. 
 
\subsection{Exact bisimulation relations for models with continuous state spaces} 
The early notion of 
bisimulation between labelled Markov chains \cite{larsen1991bisimulation} has been extended to processes (again denoted as LMPs) defined over analytical state spaces in \cite{Desharnais2002}, by employing 
zigzag morphisms. 
This work combines and extends earlier results on zigzag-based bisimulations \cite{Blute1997,Desharnais1998,Edalat1999a},  
provides the fundamental measure theoretical results to support bisimulations over continuous spaces, 
and shows their logical characterisation and their transitivity property.  
Alternative but equivalent to the zigzag definition, 
the follow-up work in 
\cite{Desharnais2003} discusses an extension of the relational notion in \cite{larsen1991bisimulation}, 
based on the concept of measurable $\mathcal R_{eq}$-closed sets. 

Suppose that we have a LMP $\mathbf S=(\mathbb X,\mathcal B(\mathbb X ),\mathbb T_l,L)$, 
with a finite label set $l\in L$ and with $\X$ being a Polish space. 
Note that, unlike in the discrete-space case, this process is defined together with a Borel $\sigma$-algebra $\mathcal B(\mathbb X )$. 
 Then based on \cite{Desharnais2003} an equivalence relation, denoted $\rel_{eq}$, defines a bisimulation if for any $x_1 \mathcal R_{eq} x_2$ and for any measurable $\mathcal R_{eq}$-closed set $B$ (or equivalently for every measurable set $B\subset \mathbb X /\rel_{eq}$) it holds that
\begin{align*}\mathbb T_{l}(B|x_1)=\mathbb T_{l}(B|x_2),\  \forall l\in L.\end{align*}

\begin{figure}[htp] 
\begin{subfigure}[b]{.48\textwidth} 
   \centering{\includegraphics[width=2.25in]{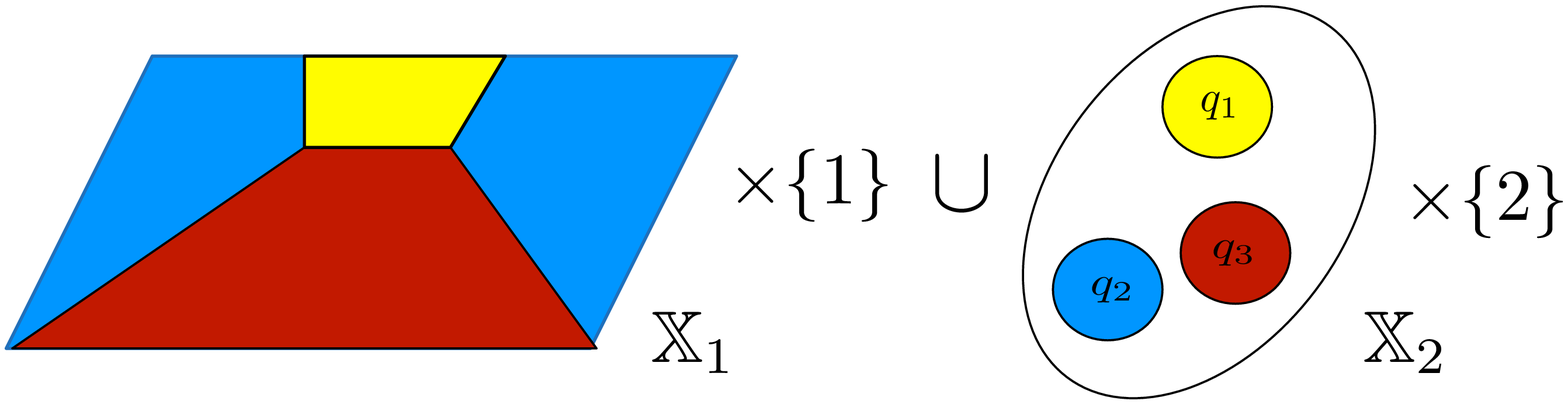} } 
   \caption{ 
An equivalence relation $\rel_{eq}$ over the disjoint union $\X_1\sqcup \X_2$, 
 where two elements from each set are in the relation if they share the same colour.  
} 
   \label{fig:disjointunion}
\end{subfigure}\hfill
\begin{subfigure}[b]{.48\textwidth}%
   \centering 
   \includegraphics[width=1.5in]{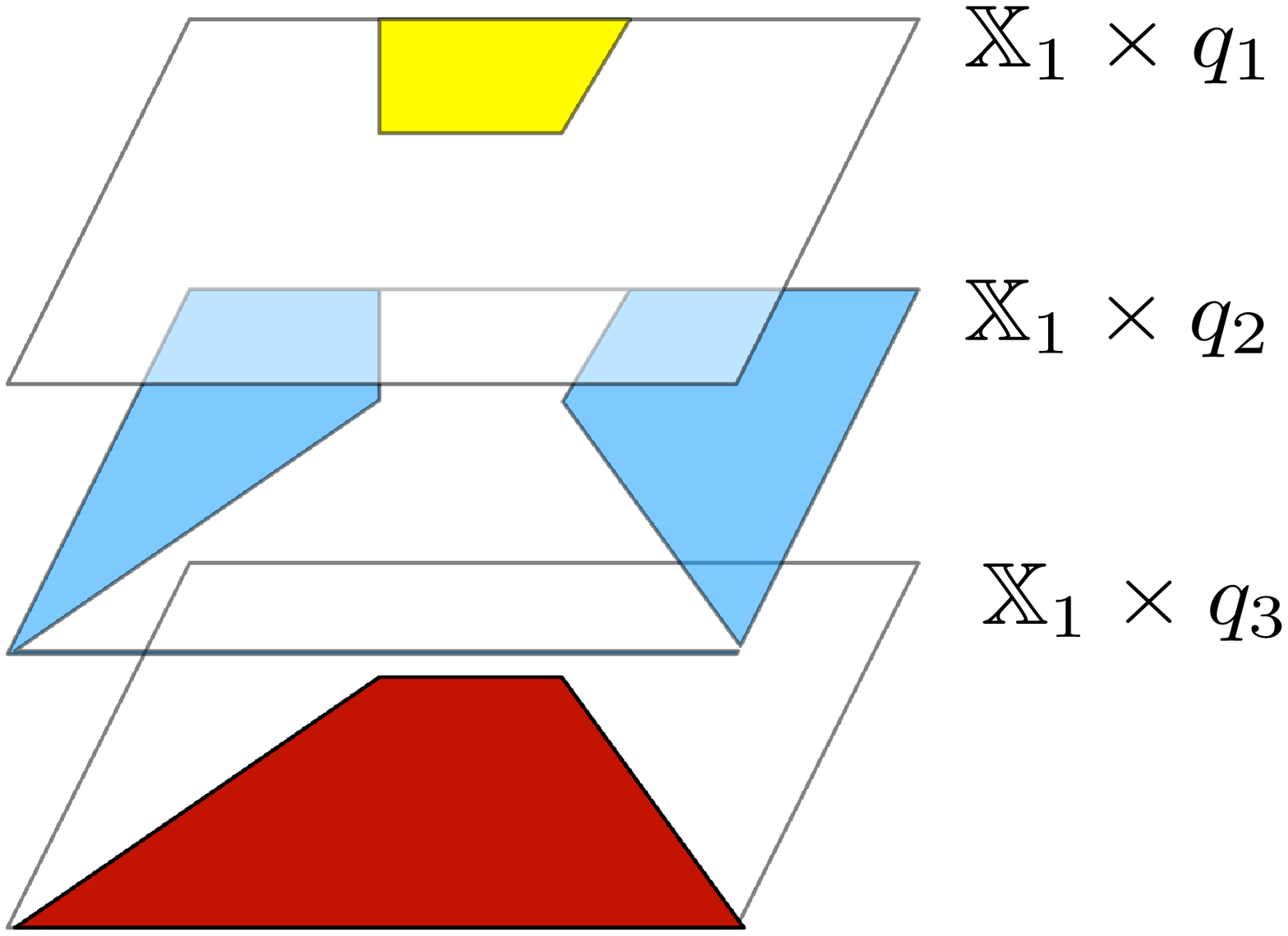} 
   \caption{Relation $\rel$ over the Cartesian product of $\X_1 \subset \mathbb R^2$ and $\X_2=\{q_1,q_2,q_3\}$, 
   induced by the equivalence relation $\rel_{eq}$. 
   Elements of the relation are coloured.}
   \label{fig:cartesian}
   \end{subfigure}\vskip-.5cm
   \caption{Example of an equivalence relation over the disjoint union of two heterogeneous spaces, and the corresponding relation over their Cartesian product. 
 }\vskip-.5cm \end{figure}

As an extension, a bisimulation between two different LMPs $\mathbf S_i=(\mathbb X_i,\mathcal B(\mathbb X_i ),\mathbb T_{l,i},L)$, $i=1,2$ 
can be constructed by working on the disjoint union of their state spaces. More precisely,   
an equivalence relation $\mathcal R_{eq}$ over $\X_1\sqcup \X_2$ defines a bisimulation if for every $x_1\rel_{eq}x_2$ (where $x_1\in \X_1$ and $x_2\in\X_2$) 
and for every $\mathcal R_{eq}$-closed set $B$, 
it holds that 
\[\mathbb T_{l,1}(B\cap \X_1|x_1)=\mathbb T_{l,2}(B\cap \X_2 |x_2),\  \forall l\in L.\]
An example of an equivalence relation over the disjoint union between two heterogeneous spaces, 
along with the induced quotient space, is given in Fig. \ref{fig:disjointunion}.  
The discussed notion of equivalence between LMPs crucially depends on the equivalence of the \emph{probability spaces} $(\mathbb X_i,\mathcal B(\mathbb X_i ),\mathbb P_i)$ with probability measures $\mathbb P_i:=\mathbb T_{l,i}(\cdot\mid x_i)$, given for a fixed $l$ and state $x_i$. 
For an equivalence relation $\rel_{eq}$ over $\X_1\sqcup\X_2$, 
the probability spaces are equivalent if for every measurable $\rel_{eq}$-closed set $B$ it holds that 
\[\mathbb P_1(B\cap \X_1)=\mathbb P_2 (B\cap \X_2),\] 
which is denoted as $\mathbb P_1\equiv_{\rel_{eq}}\mathbb P_2$. 

This type of equivalence between \emph{probability spaces} has also been used for bisimulation relations between 
control Markov processes 
\cite{Abate2011}, a simpler instance of the gMDP framework discussed in this work.  
As such, it is a natural extension of the notion in \cite{Desharnais2002,Desharnais2003} from LMPs to control Markov processes. 

An equivalence relation defined over the disjoint union of $\X_1$, and $\X_2$, i.e., $\rel_{eq}\subset (\X_1 \sqcup\X_2)\times (\X_1 \sqcup\X_2)$, 
can also be expressed as a relation over their Cartesian product, namely $\rel:=\{(x_1,x_2)\in \X_1\times\X_2: (x_1,x_2)\in\rel_{eq} \}$. 
As an example, we provide in Fig. \ref{fig:cartesian} the relation over the Cartesian product of two spaces, 
corresponding to the equivalence relation defined in Fig. \ref{fig:disjointunion} over their disjoint union.  
This connection raises the question of whether probability spaces related via $\rel_{eq}$ are also in a lifted relation. 
When working with finite or countable sets, we know that this connection holds \cite{Segala1995}.  
On the other hand, for continuous or uncountable spaces this depends on the absence of measure-theoretical issues, 
and will be studied in depth to answer when the following claim holds. 
\begin{claim} \label{thm:equivprop}
Consider two measure  spaces $(\X_1,\mathcal B (\X_1))$ and $(\X_2,\mathcal B(\X_2))$ and an equivalence relation $\rel_{eq}$ that induces a relation over $\X_1\times \X_2$ as $\rel:=\{(x_1,x_2)\in \X_1\times \X_2: (x_1,x_2)\in\rel_{eq} \}$. 
Then, 
\begin{itemize}
\item for any two probability measures $\Delta\in\mathcal{P}(\X_1,\mathcal B (\X_1))$ and  $\Theta\in\mathcal{P}(\X_2,\mathcal B (\X_2))$, we have 
\[
\Delta\bar \rel \Theta \, \textmd{ if and only if } \, \Delta\equiv_{\rel_{eq}} \Theta. 
\]
\item for any two universally measurable transition kernels $\mathbb T_1$ and $\mathbb T_2$, there exists a universally measurable kernel $\Wt$ that lifts  the transition kernels for $\rel$ as required in Def. \ref{def:pbsim}. 

\end{itemize}
 \end{claim} 

In order to prove this claim and to construct the lifted measure based on an equivalence relation, 
we exploit the notion of zigzag morphism \cite{Desharnais2002,Edalat1999a} and its properties.  

\begin{figure}[htp]
\begin{subfigure}[b]{.22\textwidth}
\resizebox{\linewidth}{!}{\begin{tikzpicture} 
\node (s1){$\mathbf S_1\!$};
\node[above right of= s1,node distance=1.2cm] (t) {$\mathbf T$};
\node[below right of= t,node distance=1.2cm] (s2) {$\mathbf S_2$};
\draw[->] (t)to node[above left]{$f$} (s1);
\draw[->] (t)to node[above right]{$g$} (s2);
\end{tikzpicture}}\caption{Generalised span of zigzag morphisms}
\end{subfigure}\hfill\begin{subfigure}[b]{.4\textwidth}
\resizebox{\linewidth}{!}{\begin{tikzpicture} 
\node[node distance=2.4cm](s1){$\mathbf S_1\!\!\!$};
\node[above right of= s1,node distance=1.3cm] (t) {$\mathbf T_{12}\!\!\!$};
\node[below right of= t,node distance=1.3cm] (s2) {$\mathbf S_2\!\!\!$};
\node[above right of= s2,node distance=1.3cm] (t2) {$\mathbf T_{23}$};
\node[above right of= t,node distance=1.3cm] (tt) {$\mathbf T^\ast$};
\node[below right of= t2,node distance=1.3cm] (s3) {$\mathbf S_3$};
\draw[->] (t)to node[above left]{$f_1$} (s1);
\draw[->] (t)to node[above right,xshift=-.1cm]{$f_2$} (s2);
\draw[->] (t2)to node[above left,xshift=.1cm]{$f_3$} (s2);
\draw[->] (t2)to node[above right]{$f_4$} (s3);
\draw[->,dashed] (tt)to node[above left]{$g_1$} (t);
\draw[->,dashed] (tt)to node[above right]{$g_2$} (t2);
\end{tikzpicture}}\caption{ Construct $\mathbf T^\ast$ as a semi-pullback of co-span $\mathbf T_{12}\rightarrow \mathbf S_2\leftarrow \mathbf T_{22}$.}\end{subfigure}\hfill
\begin{subfigure}[b]{.29\textwidth}
\resizebox{\linewidth}{!}{\begin{tikzpicture} 
\node[node distance=2.4cm] (s1){$\mathbf S_1\!$};
\node[above right of= s1,node distance=1.2cm] (t) {$\mathbf T^\ast$};
\node[below right of= t,node distance=1.2cm] (s2) {$\mathbf S_3$};
\draw[->,dashed] (t)to node[above left]{$f_1\!\circ\! g_1$} (s1);
\draw[->,dashed] (t)to node[above right]{$f_4\!\circ\! g_3$} (s2);
\end{tikzpicture}}\caption{Transitive bisimulation based on semi-pullback}
\end{subfigure}\caption{Probabilistic bisimulation between $\mathbf S_1$ and $\mathbf S_2$ established by zigzag morphism. Transitivity of probabilistic bisimulations $\mathbf S_1$ and $\mathbf S_2$ and $\mathbf S_2$ and $\mathbf S_3$ follows as a semi-pullback.}
\label{fig:ziggyzag}
\end{figure}
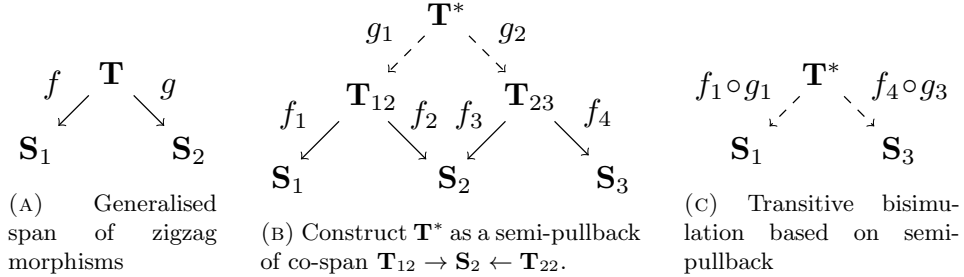

More precisely, 
consider a tuple $(\X,\mathcal{B}(\X),\mathbb T)$, 
with $\X$ a Polish space and $\mathbb T:\X\times \mathcal{B}(\X)\rightarrow [0,1]$ a transition probability function. 
\begin{defn}[Morphism] 
A function $f:(\X,\mathcal{B}(\X),\mathbb T)\rightarrow (\X',\mathcal{B}(\X'),\mathbb T')$ is a \emph{morphism} if it is a continuous surjective map $f:\X \rightarrow \X'$, 
such that for all $s\in \X$ and for all $B\in \mathcal{B}(\X)$,  
\[\mathbb T(f^{-1}(B)|s)=\mathbb T'(B|f(s)),\] 
i.e., it is preserving transition probabilities. 
\end{defn}
Consider two labelled Markov processes  $\mathbf S_i=(\X_i,\mathcal{B}(\X_i),\{k_{l,i}| l\in L\})$ with a shared finite set of labels $L$, 
then a morphism $f$ is a \emph{zigzag morphism} if it preserves the two transition probability functions for all $l\in L$. 
Two LMPs $\mathbf S_1$ and $\mathbf S_2$ are \emph{probabilistically bisimilar} if there is a generalised span of zigzag morphisms between them \cite{Desharnais2002}; 
namely, if there exists a labelled Markov process $\mathbf T$  (with universally measurable transition kernels) and zigzag morphisms $f$ and $g$ from $\mathbf T$ to $\mathbf S_1$ and $\mathbf S_2$, respectively (see Figure \ref{fig:ziggyzag}a).
In order to prove that this notion of probabilistic bisimulation is transitive, 
\cite{Edalat1999a} has shown that
\begin{itemize}
\item the category of Markov processes with universally measurable transition probability functions $\mathbb T$ on Polish spaces and with surjective and continuous transition probability preserving maps has \emph{semi-pullbacks} \cite [Corollary 5.3]{Edalat1999a}; 
\item the category of probability measures $\mathbb P$ on Polish spaces and measure-preserving surjective maps has  \emph{semi-pullbacks} \cite[Corollary 5.4]{Edalat1999a}. 
\end{itemize}
By adding a labelling to the transition probability function $\mathbb T$, one can trivially show the existence of semi-pullbacks on an LMP.
Moreover, the transitivity of probabilistic bisimulations follows based on semi-pullbacks:  if $\mathbf S_1$ is probabilistically  bisimilar to $\mathbf S_2$, which is also bisimilar to $\mathbf S_3$, then $\mathbf S_1$ and $\mathbf S_3$ are bisimilar, as in Figure \ref{fig:ziggyzag}b.

Let us go back to the Claim \ref{thm:equivprop}. Firstly recall that, as depicted in Fig. \ref{fig:disjointunion}, 
an equivalence relation $\rel_{eq}$ over $ \X_1\sqcup \X_2$ induces a quotient space, denoted by $\mathcal Q:=( \X_1\sqcup \X_2)/\rel_{eq}$,  
and partitions the unionised state space by disjoint sets, 
namely $\bigcup_{q\in\mathcal Q}q= \X_1\sqcup \X_2$, and $q_1\cap q_2=\emptyset$ for $q_1\not=q_2$, $q_1,q_2\in\mathcal Q$. 
Thus starting from the Markov processes $\mathbf S_1=(\X_1, \mathcal B(\X_1),\mathbb T_1)$  and $\mathbf S_2=(\X_2, \mathcal B(\X_2),\mathbb T_2)$,  
we show that the claim holds under either of the following two conditions. 
\begin{condition}[Polish quotient space]\label{quotient:Polish}
The equivalence relation of interest $\rel_{eq}$ induces a quotient space $(\mathcal Q,\mathcal{F})$ that is Polish  and the maps from $\X_1$ and $\X_2$ to the quotient space $f_1:\X_1\rightarrow \mathcal Q$ and $f_2:\X_2\rightarrow \mathcal Q$ are measurable and surjective.  
\end{condition}
\begin{condition}[Analytic Borel quotient space]\label{quotient:analytical}
The equivalence relation of interest $\rel_{eq}$ induces a quotient space that is analytical as in \cite{Desharnais2002,Edalat1999a} and  the maps from $\X_1$ and $\X_2$ to the quotient space $f_1:\X_1\rightarrow \mathcal Q$ and $f_2:\X_2\rightarrow \mathcal Q$ are measurable and surjective.  
\end{condition}
Notice that condition \ref{quotient:Polish} implies condition \ref{quotient:analytical}, 
and further note that $f_1$ and $f_2$ are constructed based on the injection $\iota_1$ and $\iota_2$, i.e., 
$ \iota_i:\X_i\rightarrow \X_1\sqcup\X_2$ for $i=1,2$, composed with $q:\X_1\sqcup\X_2\rightarrow Q$.

Then we can construct the quotient Markov process as the tuple $\mathbf S:=(\mathcal Q,\mathcal F, \mathbb T)$ such that  $(\mathcal Q,\mathcal F)$ is a Borel measurable space with
  $\mathcal Q=(\mathbb S_1\sqcup\mathbb S_2)/\rel_{eq}$, and $\mathcal{F}$ is defined as
 $\mathcal{F}:=\{E\subset\mathcal Q : q^{-1}(E)\in\mathcal B(\mathbb S_1\sqcup\mathbb S_2)\}$.  
The stochastic transition kernel $\mathbb T$ is
 constructed as in \cite[Proof of Proposition 9.4]{Desharnais2002}. For any $B\in \mathcal F$ it holds that 
 \begin{align}
\mathbb T (B|t)= \mathbb T_1 (f_1^{-1}(B)|s) \quad \mbox{with } s\in f_1^{-1}(t)
 \end{align}   and $\mathbb T(B|\cdot)$ is Borel measurable.
 
Then $f_1$ and $f_2$ are zigzag morphisms from,  respectively, $\mathbf S_1$ and $\mathbf S_2$ to $\bf S$, 
and they form a co-span. 
Based on \cite{Edalat1999a} we now know that there exists a Markov process $\mathbf W:=\left((\X_1\times \X_2),\mathcal{B}(\X_1\times \X_2),\mathbb W\right)$, 
which is a semi-pullback, 
and where $\mathbb W$ lifts the relation over $\X_1\times\X_2$ and defines a universally measurable stochastic kernel. 
If $\bf S_1$, $\bf S_2$ and $\bf S$ have analytical Borel spaces (this includes Polish spaces) and universally measurable transition kernels, then $\mathbb W:\rel\times \mathcal{B}(\times)$ is defined  as
\begin{align}
\mathbb W\left(dx_1'\times dx_2'\mid (x_1,x_2)\right)=\int_{q'\in Q} \mathbb T_1(dx_1'\mid x_1,q') \mathbb T_2(dx_2'\mid x_2,q')  \mathbb T(dq'\mid f_1(x_1)), 
\end{align}
where $\mathbb T_i(dx_i'\mid x_i,q') $ for $i=1,2$ are universally measurable regular conditional probability distributions, 
such that for measurable subsets $X_i\subset \X_i$ and $Q\subset \mathcal Q$ it holds that 
\begin{align*}
\mathbb T_i(X_i \cap f_{1}^{-1}(Q) \mid x_i)=\int_Q \mathbb T_i(dx_i'\mid x_i,q') \mathbb T (dq'\mid f_1(x_1)).\end{align*}
The details of this reasoning follow from \cite{Edalat1999a} together with the existence proof for the regular conditional probability distributions.

\begin{remark}[Measurability assumptions]
The measurability assumption above is a nontrivial but natural assumption, since, as proven for LMPs, 
any equivalence relation on $ \X_1\sqcup \X_2$ based on logics induces a quotient LMP that has an analytical Borel space and measurable canonical maps \cite[Proposition 9.4]{Desharnais2002}.  
\end{remark}

\subsection{Approximate probabilistic bisimulation relations}
A relaxation of exact equivalence relations in a probabilistic context has been first introduced for (finite-state) labelled Markov chains in \cite{desharnais2004metrics}, 
and later employed in \cite{cDAK12}.  
\begin{defn}\label{def:APBDesh} 
A relation $\mathcal R\subseteq S\times S$ is an (probabilistic) $\epsilon$-simulation if whenever $s\mathcal R t$, then for all labels $l\in L$, 
and sets in the event space $X\in\Sigma$, it holds that
\[\mathbb{T}_l(\mathcal{R}(X)|t)\geq \mathbb T_l (X|s)-\epsilon.\]
\end{defn}
Note that the  relation is not required to be an equivalence relation, hence it does not induce a partitioning of the state space.  
For continuous-space systems, \cite{Abate2011} has discussed an approximate (bi-)simulation notion derived from the finite-state definition. 
This definition relates to an approximate equivalence of the \emph{probability spaces} $(\mathbb X_i,\mathcal B(\mathbb X_i ),\mathbb P_i)$ $i=1,2$ as follows. 
For an equivalence relation $\rel_{eq}$ over $\X_1\sqcup\X_2$  the probability spaces are approximately equivalent if for every measurable $\rel_{eq}$-closed set $B$ it holds that
\[|\mathbb P_1(B\cap \X_1)-\mathbb P_2 (B\cap \X_2)|\leq \delta,\] 
which is denoted as $\mathbb P_1\equiv_{\rel_{eq}}^\delta\mathbb P_2$.

\begin{theorem}\label{thm:equivlifting}
Consider two measure spaces $(\X_1,\mathcal B (\X_1))$ an $(\X_2,\mathcal B(\X_2))$ and an equivalence relation $\rel_{eq}$ satisfying condition \ref{quotient:Polish}. 
Then for any two probability measures $\Delta\in\mathcal{P}(\X_1,\mathcal B (\X_1))$ and  $\Theta\in\mathcal{P}(\X_2,\mathcal B (\X_2))$ we have that 
\[\Delta\equiv^\delta_{\rel_{eq}} \Theta\textmd{ if and only if } \Delta\bar \rel_{\delta} \Theta, \]
with as standard $\rel:=\{(x_1,x_2)\in \X_1\times \X_2: (x_1,x_2)\in\rel_{eq} \}$.  
\end{theorem}
 \proof
 \noindent$\bf1.$
 $\Delta\bar \rel_{\delta} \Theta
\implies  \Delta\equiv^\delta_{\rel_{eq}} \Theta$\\*
 If  $\Delta \bar\rel_\delta \Theta$ then for each  $C\subset (\X_1\sqcup \X_2)/\rel_{eq}$ with subsets $\tilde S=\X_1\cap C\in \mathcal{B}(\X_1)$ and $\tilde T=\X_2\cap C\in \mathcal{B}(\X_2)$, then $|\Delta(\tilde S)-\Theta(\tilde T)|\leq \delta$ because
\(\mathbb
W(\tilde S\times (\X_2\setminus \tilde T))\leq \delta 
\)
and 
\(\mathbb W((\X_1\setminus \tilde S)\times \tilde T)\leq \delta\). This can be shown as follows 
\[ \Delta(\tilde S)\leq \Delta(\tilde S)+\mathbb W((\X_1\setminus \tilde S)\times\tilde T)=\Theta(\tilde T)+\mathbb W(\tilde S\times (\X_2\setminus \tilde T))\leq \Theta(\tilde T)+\delta \]
and repeating the reasoning starting from $\Theta(\tilde T)$ we get $\Theta(\tilde T)\leq \Delta(\tilde S)+\delta$, and hence $|\Delta(\tilde S)-\Theta(\tilde T)|\leq \delta$.\\*
\noindent$\bf2.$
 $\Delta\equiv^\delta_{\rel_{eq}} \Theta\implies \Delta\bar \rel_{\delta} \Theta$\\*
 Under Condition \ref{quotient:Polish} we have that the quotient space has the Borel measure space $(\mathcal Q, \mathcal F)$ where $\mathcal Q$ is Polish.
Additionally we have measurable mappings $f_i:\X_1\rightarrow \mathcal Q$. We denote the induced probability measures ${f_1}_\ast \Delta\in \mathcal P(\mathcal Q, \mathcal F)$ and ${f_2}_\ast \Theta\in \mathcal P(\mathcal Q, \mathcal F)$. 
Denote a measure that lifts these over the diagonal relation as $\mathbb W_{\mathcal Q}\in\mathcal P(\mathcal Q^2 , \mathcal F^2) $. This is equivalent to maximal coupling of ${f_1}_\ast \Delta$ and ${f_2}_\ast \Theta$. Specifically for Polish spaces we  take the $\gamma$-coupling given as $\mathbb W_{\mathcal Q} := \gamma({f_1}_\ast \Delta,{f_2}_\ast \Theta) \in\mathcal P(\mathcal Q^2 , \mathcal F^2)$ 
\cite{art2014}
based on \cite[Section 1.5]{lindvall2002lectures} and given as follows
\begin{definition}
Let $Z$ be a Borel space and let $\nu, \tilde\nu \in\mathcal(Z)$ be two probability measures on it. The $\gamma$-coupling of $(\nu, \tilde\nu)$ is a measure $\gamma\in\mathcal(Z^2)$ given by
\begin{align*}
\gamma(\nu, \tilde\nu):=\Psi_Z(\nu\wedge \tilde\nu)+\mathbf{1}_{[0,1)} (\|\nu\wedge \tilde\nu\|).\frac{(\nu- \tilde\nu)^+\otimes(\nu- \tilde\nu)^-}{1-\|\nu-\tilde\nu\|}
\end{align*}
where $\Psi_Z:Z\rightarrow Z^2$ is the diagonal map on $Z$ given by $\Psi_Z:z \mapsto (z,z)$.
\end{definition} 
\noindent The lifted measure over $\mathbb W\in\mathcal P(\X_1\times\X_2,\mathcal B(\X_1\times\X_2))$
is given as 
\[\mathbb W:= \int_{Q\times Q} \Delta(dx_1\mid q_1) \Theta(dx_2\mid q_2)  \mathbb W_{\mathcal Q} (dq_1\times d q_2).\]
 \endproof

 \section{Proofs of Theorems and Corollaries}\label{app:proofs} 
 
 \subsection{Control refinement proofs, Theorem 1- 4}
 Let us consider the controller refinement for exact simulation relations first.
 The execution $\{( x_2(t), x_{\C_2}(t)){\mid} t\in [0,N]\}$, 
is defined on the canonical space $\Omega=(\X_2\times \X_{\C_2})^{N+1}$, 
and has a unique probability measure $\mathbb P_{\C_2\times \M_2}$. 
Therefore in Alg. 1, 
in order to write the execution of the refined control $\C_2$ and of the gMDP $\M_2$, 
we have included the state of $\M_2$ for one transition in the state of the refined control strategy. 
Therefore, while the execution of Alg. 1 ranges over $\X_{\C_1}\times\X_1\times \X_2$, 
the execution of the controlled system with $\C_2$ ranges over $\X_{\C_2}\times \X_{2}=(\X_{\C_1}\times\X_1\times \X_2)\times\X_2$. The marginal of $\mathbb P_{\C_2\times \M_2}$ on $\X_{\C_1}\times\X_1\times \X_2$ defines the measure for the execution in Alg.1.

Since, by the above construction of $\C_2$, 
the output spaces of the closed loop systems $\C_1\times\M_1$ and $\C_2\times \M_2$ have equal distribution, 
it follows that measurable events have equal probability, as stated next. 

 \begin{proof}[of Theorem \ref{thm:events}]
If $\{h_1(x_1(t)){\mid}t\in[0,N]\}\in A$ and $(x_1(t),x_2(t))\in\rel$ $\forall t\in[0,N]$ then $\{h_2(x_2(t)){\mid}t\in[0,N]\}\in A$.\\ \medskip

Let us rewrite the stochastic kernel of the combined transition of $\C_2$ and $\M_2$ for $t=0$ as\footnote{For brevity a part of the argument of the stochastic kernel has been omitted.} 
\[
\mathbb T^0_{\C_2\times \M_2}(d x_{\C_2}\times dx_2) = 
\mathbb T^0_{\C_1} (dx_{\C_1}{\mid}x_{\C_10},x_1)\mathbb W_\pi(dx_1 {\mid} x_2 )\delta_{x_2(0)}(dx_2)\pi(dx_2(0)). 
\]
Marginalised on $\X_{\C_1}\times \X_1\times\X_2$, this becomes (by definition of $\mathbb W_{\pi}$)
\begin{align*}\mathbb T^0_{\C_2\times \M_2}(d x_{\C_1}\times dx_1\times dx_2)&= 
\mathbb T^0_{\C_1} (dx_{\C_1}{\mid}x_{\C_10},x_1)\mathbb W_\pi(dx_1 {\mid} x_2 )\pi(dx_2)\\
&=\mathbb T^0_{\C_1} (dx_{\C_1}{\mid}x_{\C_10},x_1)\mathbb W_\pi(dx_2 {\mid} x_1 )\pi(dx_1). 
\end{align*}
Further marginalised on $\X_{\C_1}\times\X_1$, this becomes 
\begin{align*}
\mathbb T^0_{\C_2\times \M_2}(d x_{\C_1}\times dx_1)= 
\mathbb T^0_{\C_1} (dx_{\C_1}{\mid}x_{\C_10},x_1)\pi(dx_1)= \mathbb T^0_{\C_1\times \M_1}(d x_{\C_1}\times dx_1).
\end{align*}
For $t\in [1, N]$, 
the stochastic kernel marginalised on $\X_{\C_1}\times\X_1\times \X_2$ is 
\begin{align*}&\mathbb T^t_{\C_2\times \M_2}(d x_{\C_1}'\times dx_1'\times dx'_2)= \mathbb T^t_{\C_2} (dx_{\C_1}'{\mid}x_{\C_1},x_1')\\ &\hspace{3cm} \Wt(d x_1' {\mid} x_2',h_{\C_1}^t(x_{\C_1}),x_2, x_1)\mathbb T_2(dx_2'{\mid}x_2,h^t_{\C_2}(x_{\C_2}) )\\
&\hspace{3cm}= \mathbb T^t_{\C_1} (dx_{\C_1}'{\mid}x_{\C_1},x_1')\Wt(d x_1'\times d x_2' {\mid} h_{\C_1}^t(x_{\C_1}),x_2, x_1)\end{align*}
and can be further marginalised on $\X_{\C_1}\times \X_1$ to obtain $\mathbb T^t_{\C_1\times \M_1} $. 
Note that since $\Wt(\rel {\mid} h_{\C_1}^t(x_{\C_1}),x_2, x_1)=1$ for $(x_1,x_2)\in \rel$ it holds with probability $1$ that $(x_1(t),x_2(t))\in \rel$ for $t\in[0,N]$. 
Therefore we can deduce that 
 \[ \po_{\C_1\times\M_1}\left(\{y_1 (t)\}_{0:N}\in A\right)= \po_{\C_2\times\M_2}\left(\{y_2 (t)\}_{0:N}\in A\right).\] 
\qed \end{proof}

To prove Theorem \ref{thm:Apprxstrat} and \ref{thm:cr} we leverage their exact versions (Theorem \ref{thm:Cs1}
 and \ref{thm:events}).
 We first show the existence of a refined control strategy in case of approximate simulation relation, c.f. Theorem \ref{thm:Apprxstrat}.
 Then we leverage these results to prove Theorem  \ref{thm:cr}.
 
 Theorem \ref{thm:Apprxstrat} states the following.
Let gMDP $\M_1$ and $\M_2$, with $\M_1\preceq_\eps^\delta\M_2$, and control strategy $\C_1=(\X_{\C_1},x_{\C_10},\X_1,\mathbb T_{\C_1}^t,h_{\C_1}^t)$ for $ \M_1$ be given. 
Then for every given recovery control strategy $\C_{rec}$, 
a refined control strategy $\C_2=(\X_{\C_2},x_{\C_20},\X_2,\mathbb T_{\C_2}^t,h_{\C_2}^t)$ can be obtained as an \emph{inhomogenous Markov process} with two discrete modes of operation,  
$\{\operatorname{refinement}\}$ and $\{\operatorname{recovery}\}$, based on Algorithm 2. 
More specifically a possible choice of a refined control strategy is build up as follows
\begin{itemize}
\item state space $\X_{\C_2}:=\{\X_{\C_1}\times \X_1\times \X_2\times \{\operatorname{refine}\}\}\cup \X_{\C_{rec}}\times \{\operatorname{recover}\} $ with elements $x_{\C_2}=(x_{\C_1}, x_1,x_2, \operatorname{refine})$ and $x_{\C_2}=(x_{\C_{rec}}, \operatorname{recover})$; 
\item initial state $x_{\C_20}:=(x_{\C_10},0,0,\operatorname{refinement})$;
 \item  accepting as control inputs $x_2\in\X_2$;
\item time dependent stochastic kernel $\mathbb T^t_{\C_2}$, defined for $t=0$ as
\begin{align*}
\mathbb T^0_{\C_2} (dx_{\C_2}^{\operatorname{refine}}{\mid}x_{\C_20},x_2(0) ):=&\mathbb T_{\C_1}^0 (dx_{\C_1}{\mid}x_{\C_10},x_1)\mathbf 1_{\rel}\left(x_1,x_2\right)\\&\qquad\times \mathbb W_\pi(dx_1 {\mid} x_2 )\delta_{x_2(0)}(dx_2)\\
\mathbb T^0_{\C_2} (dx_{\C_2}^{\operatorname{recover}}{\mid}x_{\C_20},x_2(0) ):=&\mathbb T^0_{init,rec} (dx_{\C_{rec}}{\mid}x_2)\mathbf 1_{(\X_1\times \X_2)\setminus\rel}\left(x_1,x_2\right)\\&\qquad\times \mathbb W_\pi(dx_1 {\mid} x_2 )\delta_{x_2(0)}(dx_2)
\end{align*}
and for $t\in[1,N]$ over the $\{\operatorname{refine}\}$ operating mode 
\begin{align*}
\mathbb T^t_{\C_2} (dx_{\C_2}^{\operatorname{refine}'}{\mid}x_{\C_2}^{\operatorname{refine}}(t),x_2(t) )&:=\mathbb T^t_{\C_1} (dx_{\C_1}'{\mid}x_{\C_1},x_1')\mathbf 1_{\rel}(x_1',x_2')\\ &\qquad\times
\Wt(d x_1' {\mid} x_2',h_{\C_1}^t(x_{\C_1}),x_2, x_1)\delta_{x_2(t)}(dx'_2);\\
\mathbb T^t_{\C_2} (dx_{\C_2}^{\operatorname{recover}'}{\mid}x_{\C_2}^{\operatorname{refine}}(t),x_2(t) )&:=\mathbb T^t_{init,rec}(dx_{\C_{rec}}'{\mid}x_2')\mathbf 1_{(\X_1\times \X_2)\setminus\rel}(x_1',x_2')\\ &\qquad\times
\Wt(d x_1' {\mid} x_2',h_{\C_1}^t(x_{\C_1}),x_2, x_1)\delta_{x_2(t)}(dx'_2);
\end{align*}
defined based on a  stochastic kernel $\mathbb T^t_{init,rec}$  $t\in[0,N]$ initiates the recovery strategy on the fly and is contained in the choice of recovery strategy. 
And for $t\in[1,N]$ for the $\operatorname{recover}$ operating mode 
\begin{align*}
\mathbb T^t_{\C_2} (dx_{\C_2}^{\operatorname{recover}'}{\mid}x_{\C_2}^{\operatorname{recover}}(t),x_2(t) )&:=\mathbb T^t_{\C_{rec}} (dx_{\C_{rec}}'{\mid}x_{\C_{rec}}(t),x_2(t) );
\end{align*}

\item universally  measurable output map \[h^t_{\C_2}(x_{\C_2}):= \left\{\begin{array}{ll}\InF(h^t_{\C_1}(x_{\C_1}),x_1, x_2) &\quad \mbox{ for $\operatorname{refine}$\,,}\\
h^t_{\C_{rec}}(x_{\C_{rec}}) &\quad \mbox{ for $\operatorname{recover}$\,.}\end{array}\right.\, \] 
 
\end{itemize}
The refined control strategy is composed of the control strategy $\C_1$, the recovery strategy $\C_{rec}$, the stochastic kernel $\Wt$, and the interface $\InF$.  
Both the time-dependent stochastic kernels $\mathbb T^t_{\C_2}$ and the output maps $h_{\C_2}^t$, for $t\in[0,N]$, can be shown to be universally measurable, 
since Borel measurable maps (and kernels) are universally measurable and the latter are closed under composition \cite[Ch.7]{bible}.  

Now we need to use this control strategy to prove Theorem \ref{thm:cr}.

\begin{proof}[of Theorem \ref{thm:cr}]
Given $\C_{rec}$ consider an auxiliary recover strategy $\C_{rec}^\ast$ such that it has stochastic kernels over $\X_{\C_{rec}}\times \X_1\times \X_{\C_1}$:
\begin{align*}\mathbb T^t_{\C_{rec}^\ast} (dx_{\C_{rec}^\ast}'{\mid}x_{\C_{rec}^\ast}(t),x_2(t) )&=\mathbb T^t_{\C_{rec}} (dx_{\C_{rec}}'{\mid}x_{\C_{rec}}(t),x_2(t) )\\&\hspace{2cm }\mathbb T^t_{\C_{1}\times \M_1} (dx_{\C_{1}\times \M_1}'{\mid}x_{\C_{1}\times \M_1}(t)) \end{align*}
where $\mathbb T^t_{\C_{1}\times \M_1} (dx_{\C_{1}\times \M_1}'{\mid}x_{\C_{1}\times \M_1}(t)$ is the stochastic kernel over   $\X_{\C_1\times \M_1}:=\X_1\times \X_{\C_1}$.
Due to the independence of this kernel the probability distribution $\mathbb P_{\C_2^\ast\times \M_2}$ of $\M_2$ controlled by $\C_2^\ast$ is, when marginalised on the canonical sample space $(\X_{\C_2}\times \X_{\M_2})^{N+1}$, equal to $\mathbb P_{\C_2\times \M_2}$.

Now using the same arguments as in the proof of Theorem \ref{thm:events} we know that for all measurable sets $L\subset \Y^{N+1}$
\begin{align*}
\mathbb P_{\C_1\times \M_1}(\{h_1(x_1(t))\}_{0:N}\in L)=\mathbb P_{\C_2^\ast\times \M_2}(\{h_1(x_1(t))\}_{0:N}\in L).
\end{align*}
The probability 
\[\mathbb P_{\C_2^\ast\times \M_2}\left((x_1(t),x_2(t))\in \rel \mbox{ for } t\in[0,N]\right)\geq (1-\delta)^{N+1}.
\]
This can be shown by induction starting from $t=0$, and by showing that at every time step and for every pair of states the probability of staying in $\rel$ is at least $1-\delta$.
Now note that if $\{h_1(x_1(t))\}\in A_{-\eps}$ and $(x_1(t),x_2(t))\in \rel \mbox{ for } t\in[0,N]$ then $\{y(t)\}_{0:N}\in A$.
As a consequence
\begin{align*}
 &\mathbb P_{\C_2^\ast\times \M_2}(\{h_1(x_1(t))\}_{0:N}\in A_{-\eps}\wedge (x_1(t),x(t))\in \rel \mbox{ for } t\in [0,N] )\\&\qquad\leq  \mathbb P_{\C_2^\ast\times \M_2}(\{h_2(x_2(t))\}_{0:N}\in A)= \mathbb P_{\C_2\times \M_2}(\{h_2(x_2(t))\}_{0:N}\in A). 
\end{align*}
Now using the union bounding argument we also have that
\begin{align*}
&\mathbb P_{\C_2^\ast\times \M_2}(\{h_1(x_1(t))\}_{0:N}\in A_{-\eps})-(1-\delta)^{N+1}\\
&\quad\leq
\mathbb P_{\C_2^\ast\times \M_2}(\{h_1(x_1(t))\}_{0:N}\in A_{-\eps}\wedge (x_1(t),x(t))\in \rel \mbox{ for } t\in [0,N] )
\end{align*}

 \begin{align*}&1-\mathbb P_{\C_2^\ast\times \M_2}(\{h_1(x_1(t))\}_{0:N}\in A_{-\eps}\wedge (x_1(t),x(t))\in \rel \mbox{ for } t\in [0,N] )\\
 &\leq (1-\mathbb P_{\C_2^\ast\times \M_2}(\{h_1(x_1(t))\}_{0:N}\in A_{-\eps} ))\\ &\hspace{2cm}
 + (1-\mathbb P_{\C_2^\ast\times \M_2} \left(((x_1(t),x(t))\in \rel \mbox{ for } t\in [0,N] )\right)\\
  &\leq (1-\mathbb P_{\C_2^\ast\times \M_2}(\{h_1(x_1(t))\}_{0:N}\in A_{-\eps} )) + (1-(1-\delta)^{N+1}). 
\end{align*}
We have deduced that 
\begin{align*}
\mathbb P_{\C_1\times \M_1}(\{h_1(x_1(t))\}_{0:N}\in A_{-\eps} )-(1-(1-\delta)^{N+1}) \leq \mathbb P_{\C_2\times \M_2}(\{h_2(x_2(t))\}_{0:N}\in A). 
\end{align*}
If $\{ h_2(x_2(t))\}_{0:N}\in A$ and $(\tilde x(t),x(t))\in \rel $ then $\{h_1(x_1(t))\}_{0:N}\in A_{\eps}$. Thus via similar arguments it can be deduced that 
\begin{align*}
\mathbb P_{\C_2\times \M_2}(\{h_2(x_2(t))\}_{0:N}\in A)
\leq 
\mathbb P_{\C_1\times \M_1}(\{h_1(x_1(t))\}_{0:N}\in A_{\eps} )+(1-(1-\delta)^{N+1}). 
\end{align*} 
\mbox{ }\qed
 \end{proof}

\subsection{Proof of transitivity statements}
\begin{proof}[of Theorem \ref{thm:prop} and Corollary \ref{cor:prop}]
 Since $\M_1\preceq^{\delta_a}_{\epsilon_a}\M_2$ and  $\M_2\preceq^{\delta_b}_{\epsilon_b}\M_3$
there exist
\begin{itemize}
\item relations $\rel_{12}\subset \X_1\times \X_2$ and $\rel_{23}\subset\X_2 \times \X_3$ that satisfies the required conditions in Def. \ref{def:apbsim}.
\item Interface ${\InF}_{12}: \A_1\times \X_1\times\X_2 \rightarrow \mathcal{P}(\A_2,\mathcal B(\A_2)),$ and  ${\InF}_{23}: \A_2\times \X_2\times\X_3 \rightarrow \mathcal{P}(\A_3,\mathcal B(\A_3)),$
\item and corresponding stochastic kernels ${\Wt}_{12}$ and ${\Wt}_{23}$.\end{itemize}

Define the relation $\rel_{13}\subset \X_1\times\X_3$ as $\rel_{13}:=\{(x_1,x_3)\in\X_1\times\X_3\mid\exists x_2\in\X_2 :(x_1,x_2)\in\rel_{12}, (x_2,x_3)\in\rel_{23}\}$. 
Then 
$\forall (x_1,x_3)\in \rel_{13}$ there exists a $ x_2 \in \X_2 :(x_1,x_2)\in\rel_{12}, (x_2,x_3)\in\rel_{23}$.
More specifically define a Borel-measurable function $F:\X_1\times\X_3\rightarrow\X_2$ such that $\forall (x_1,x_3)\in \rel_{13}$ for the mapping $x_2=F(x_1,x_3)$ it holds that $(x_1,x_2)\in\rel_{12}, (x_2,x_3)\in\rel_{23}$.

We have $\forall (x_1,x_3)\in\rel_{13}$ and  $x_2=F(x_1,x_3)$ :
\begin{enumerate}
\item $\cramped{\mathbf{d}\left(h_1(x_1(t)),h_3(x_3)\right)\leq \mathbf{d}\left(h_1(x_1(t)),h_2(x_2(t))\right)+\mathbf{d}\left(h_2(x_2(t)),h_3(x_3)\right)\leq \epsilon_a+\epsilon_b}$;
\item \(\cramped{\forall u_1\in\A_1:
\mathbb T_1(\cdot| x_1, u_1)\ \bar \rel_{12,\delta_a} \  \mathbb T_2(\cdot| x_2, {\InF}_{12}(u_1,x_1,x_2))}\)
and for all $u_2\in\A_2:$ 
\(\mathbb T_2(\cdot| x_2, u_2)\ \bar \rel_{23,\delta_b} \  \mathbb T_3(\cdot| x_3, {\InF}_{23}(u_2,x_2,x_3))\) 
and 
${\Wt}_{23}\in\mathcal{P}(\X_2\times\X_3,\mathcal{B}(\X_2\times\X_3))$ lifted with ${\Wt}_{12}(\cdot| u_1,x_1,x_2)$ and ${\Wt}_{23}(\cdot| u_2,x_2,x_3)$. 
\end{enumerate}

Let us derive the stochastic kernel ${\Wt}_{13}$ by combining ${\Wt}_{12}$ and ${\Wt}_{23}$ and marginalising over $\X_2$
\begin{align*}{\Wt}_{13}(dx_1'\times dx_3'|u_1,x_1,x_2,x_3)&= \int_{\X_2}{\Wt}_{23}(dx_3'\mid x_2',\InF(u_1,x_1,x_2),x_2,x_3)\\&\hspace{3cm}\times  {\Wt}_{12}(dx_1'\times dx_2'|u_1,x_1,x_2). \end{align*} 
Composed with the mapping $F$ we get a Borel-measurable stochastic kernel ${\Wt}_{13}(dx_1'\times dx_3'|u_1,x_1,x_3):={\Wt}_{13}(dx_1'\times dx_3'|x_1,F(x_1,x_3),x_3)$. In the sequel we drop the argument of the stochastic kernel.
Note that $\mathbb T_2(dx_2| x_2, \mu_{u,2})={\Wt}_{12}(\X_1 \times dx_2)={\Wt}_{23}( dx_2\times \X_3)$. 
For lifting we have to proof that ${\Wt}_{13}(\rel_{13})\geq 1-\delta_a-\delta_b$ or equivalently that ${\Wt}_{13}(\X_1\times \X_3\setminus \rel_{13})\leq \delta_a+\delta_b$, namely 
\begin{align*}&{\Wt}_{13}(\X_1\times\X_3\setminus \rel_{13})=\int_{\X_1} \int_{\X_2}\int_{\X_3\setminus \rel_{13}(x_1)}{\Wt}_{23}(dx_3\mid x_2) {\Wt}_{12}(dx_1\times dx_2)\\
&=\int_{\rel_{12}}\int_{\X_3\setminus \rel_{13}(x_1)}{\Wt}_{23}(dx_3\mid x_2) {\Wt}_{12}(dx_1\times dx_2)\\
&+\int_{\X_1} \int_{\X_2\setminus \rel_{12}(x_1)}\int_{\X_3\setminus \rel_{13}(x_1)}{\Wt}_{23}(dx_3\mid x_2) {\Wt}_{12}(dx_1\times dx_2)\\
\shortintertext{for all $(x_1,x_2)\in \rel_{12}: \rel_{23}(x_2)\subseteq\rel_{13}(x_1)$}
&\leq\int_{\X_2}\int_{\X_3\setminus \rel_{23}(x_2)}   \int_{\rel_{12}^{-1}(x_2)}{\Wt}_{12}(dx_1\mid x_2) {\Wt}_{23}(dx_2\times d x_3) \\
&+\int_{\X_1} \int_{\X_2\setminus \rel_{12}(x_1)}\int_{\X_3\setminus \rel_{13}(x_1)}{\Wt}_{23}(dx_3\mid x_2) {\Wt}_{12}(dx_1\times dx_2)\\
&\leq\int_{\X_2}\int_{\X_3\setminus \rel_{23}(x_2)}   \int_{\X_1}{\Wt}_{12}(dx_1\mid x_2) {\Wt}_{23}(dx_2\times d x_3) \\
&+\int_{\X_1} \int_{\X_2\setminus \rel_{12}(x_1)}\int_{\X_3}{\Wt}_{23}(dx_3\mid x_2) {\Wt}_{12}(dx_1\times dx_2)\\
&=\int_{\X_2}\int_{\X_3\setminus \rel_{23}(x_2)}     {\Wt}_{23}(dx_2\times d x_3)  +\int_{\X_1} \int_{\X_2\setminus \rel_{12}(x_1)} {\Wt}_{12}(dx_1\times dx_2)\\
&\leq \delta_a+\delta_b. 
\end{align*}   
In addition it has to hold that ${\Wt}_{13}(X_1\times \X_3)=\mathbb T_{1}(\cdot|x_1,\mu_{u,1})$, namely 
\begin{align*}
\mathbb {\Wt}_{13}(X_1\times \X_3)&=\int_{X_1} \int_{\X_3}\int_{\X_2}{\Wt}_{23}(dx_3\mid x_2) {\Wt}_{12}(dx_1\times dx_2)\\
&=\int_{X_1}\int_{\X_2} \int_{\X_3}{\Wt}_{23}(dx_3\mid x_2) {\Wt}_{12}(dx_1\times dx_2)\\&= {\Wt}_{12}(X_1\times \X_2)= \mathbb T_{1}(\cdot|x_1,\mu_{u,1}). 
\end{align*}
The condition ${\Wt}_{13}(\X_1\times X_3)=\mathbb T_{3}(\cdot|x_3,\mu_{u,3})
$ can be proven via similar arguments. In conclusion $\mathbb T_{1}(\cdot|x_1,\mu_{u,1})
\bar\rel_{13,\delta_a+\delta_b}\mathbb T_{3}(\cdot|x_3,\mu_{u,3})$.
To complete the proof we can show, using the same arguments as before,  that if $\pi_1\bar\rel_{12,\delta_a}\pi_2$ and if $\pi_2\bar\rel_{23,\delta_b}\pi_3$ then $\pi_1\bar\rel_{13,\delta_a+\delta_b}\pi_3$.
\qed
\end{proof}

\end{document}